\documentstyle[aps,eqsecnum]{revtex}  
\begin{document}
\tightenlines  
\draft  
\preprint{IFUM 559/FT}  
\title{Incoherent dynamics in neutron-matter interaction}  
\author{Ludovico~Lanz\footnote{E-mail: lanz@mi.infn.it}  
and Bassano~Vacchini\footnote{E-mail: vacchini@mi.infn.it}}  
\address{Dipartimento di Fisica  
dell'Universit\`a di Milano and Istituto Nazionale di Fisica  
Nucleare, Sezione di Milano, Via Celoria 16, I-20133, Milan,  
Italy}  
\date{\today}  
\maketitle  
\begin{abstract}  
Coherent and  incoherent neutron-matter interaction 
is studied inside a recently introduced  
approach to subdynamics of a macrosystem.   
The equation  
describing the interaction 
is of the Lindblad type  
and using the Fermi pseudopotential we show that the  
commutator term is an  
optical potential leading to well-known relations in neutron  
optics.   
The other terms, usually ignored in optical descriptions and   
linked  
to the dynamic structure function  
of the medium,   
give an  incoherent  
contribution to the dynamics, which keeps diffuse scattering and  
attenuation of the  coherent beam into account, thus warranting  
fulfilment of the optical theorem.   
 The relevance of this  
analysis to experiments in neutron interferometry  
is briefly  discussed.  
\end{abstract}  
\section{INTRODUCTION}  
In recent years there has been a rapidly growing interest in the  
field of particle optics, especially neutron and atom optics (for  
a recent review see~\cite{Erice,6,Sears,20,21} and~\cite{Mlynek}  
respectively,  
and references quoted therein), due to a spectacular improvement  
of the experimental techniques, connected to the introduction of  
the single crystal interferometer in the first case, and to  
progress in microfabrication technology and development of  
intense tunable lasers in the second one. Such new achievements  
provide very important tests verifying the validity of quantum  
mechanics, especially in that it predicts wavelike behaviors  
even for single microsystems.  
\par  
At the same time a new challenge arises, linked to the accuracy  
required in the description of the interaction between the  
microsystem and the apparatus acting as optical device. The  
question of the description of the dynamics of a  microsystem  
interacting with a system having many degrees of freedom (e.g.,  
matter seen as an optical medium characterized by an index of  
refraction) has been extensively studied and contains some  
typical quantum mechanical features, such as quantum  
correlations between the two systems, by which  
a reduced description  
of the  microsystem's degrees of freedom can arise only by  
suitable approximations. This subtle  
point is particularly important in the case of particle optics,  
where the main interest is devoted to the  coherent wavelike  
behavior of particles, as can be justified on the basis of the  
similarity between a Schr\"odinger equation with an optical  
potential and the Helmholtz wave equation~\cite{Sears,Mlynek}.  
The very existence of such an optical description of the  
interaction is far from trivial and strongly depends  on the  
experimental conditions. The attention has been mostly devoted to  
exploiting the optical analogies, while little has been said on  
the borderline between the optical regime, in which  coherent  
effects are predominant and a classical wavelike description  
plays a major role, and an  incoherent regime, where  incoherent  
effects, caused by the interaction between the  microsystem and  
the apparatus and showing typical particlelike features, should not be  
neglected.  
This attitude is exemplified in neutron optics by the use of  
the ``coherent wave'' formalism, instead of a reduced  
density matrix description, as usually adopted in quantum  
optics.  
\par  
In this paper we want to address the question of how to   
consistently describe      
both regimes applying a recently developed  
approach to the description of irreversible subdynamics in 
quantum mechanics~\cite{art1,japan,berlin} to the specific case  
of neutron-matter  
interaction. 
In this approach the use of an effective T-matrix describing the local  
interactions as practical starting point leads to the introduction  
of a time scale and in the particular case of particle-matter interaction 
to a dynamical  
semigroup, whose generator has the  
typical Lindblad form~\cite{Lind}. 
The expressions appearing in the generator are linked to  
particle-particle interactions, like the Fermi pseudopotential, and  
to properties of the macroscopic system, like the dynamic structure  
function, first introduced by van Hove~\cite{vanHove}. 
The  
first part of the generator  accounts for the  
description of the  coherent interaction in terms of optical  
potential and index of refraction well-known in neutron  
optics~\cite{Sears,Lax,Gold}.  
The remaining part  is shown to be related to  
the dynamic structure  
function or, equivalently, to the  density correlation function  
and leads in a  
straightforward way to results obtained in the so called  
``rigorous theory of dispersion''~\cite{Sears}.  
\par  
The paper is organized as follows: in Sec.~II we give an  
account of  
the formalism; in Sec.~III it is applied to  
neutron optics; in Sec.~IV we consider  
diffuse scattering, connection to the  
dynamic structure  
function and fulfilment of the optical theorem;  
in Sec.~V we evaluate possible  
experimental   consequences; in Sec.~VI we  
comment on our results indicating potential future developments.  
\section{INTRODUCTION OF THE FORMALISM}  
In this section we briefly introduce the formal scheme  
restricted to the description of a microsystem 
following~\cite{art1}, to which we refer the reader for  
further details.  
We indicate by ${{\cal H}^{(1)}}$ the Hilbert space in which the  
microsystem is to be described;  
its energy eigenvalues are ${E_f}$, with energy eigenstates  
$u_f$, spanning ${{\cal H}^{(1)}}$. Both systems will be considered  
confined, e.g., in a box.  
We shall adopt the second quantization formalism, setting for the  
Hamiltonian $H$ of the system:  
        \[  
        {H}={H}_0 + {H}_{\text{m}} + {V}  
        \qquad  
        \qquad  
        {H}_0 = \sum_f  
        {E_f} {a^{\scriptscriptstyle \dagger}_{f}} {a_{{f}}}  
         \qquad  
         \qquad  
        {\left[{{a_{{f}}},{a^{\scriptscriptstyle 
        \dagger}_{g}}}\right]}_{\mp}=\delta_{fg} 
        \]  
where ${a_{{f}}}$ is the destruction operator for the microsystem,  
either a Fermi or a Bose particle, in  
the state $u_f$; ${H}_{\text{m}}$ is the Hamilton operator for the  
sole  
macrosystem ($\left[{{H}_{\text{m}},{a_{{f}}}}\right]=0 $).  
Indicating by ${{{\cal H}_{\scriptscriptstyle F}}}$ the whole  
Fock space and by ${{{\cal H}^0_{\scriptscriptstyle F}}}$  
its subspace in which ${N}  =  \sum_{{h} }  
{a^{\scriptscriptstyle \dagger}_{h}}  {a_{h}}$, the number of  
microsystems, is equal to zero, we will denote  
with ${\vert \lambda \rangle}$  
the basis of eigenstates of  
${H}_{\text{m}}$ spanning ${{{\cal H}^0_{\scriptscriptstyle F}}}$,  
        $  
        {H}_{\text{m}}{\vert \lambda \rangle}=  
        E_\lambda  
        {\vert \lambda \rangle}  
        $, $N{\vert \lambda \rangle}=0$.  
${V}$~represents  
the interaction potential between the two systems.  
Having in mind to describe situations in which only one particle  
 is  
observed in each experimental run,  
or equivalently a collection of noninteracting particles in  
each run, we assume for the   statistical operator  
the following expression:  
        \[  
        {\varrho}=  
        \sum_{{g} {f}}{}  
        {a^{\scriptscriptstyle \dagger}_{g}}  
        {{\varrho}^{\text{m}}} {a_{{f}}}  
        {{\varrho}}_{gf}    ,  
        \]  
where ${{\varrho}^{\text{m}}}$ is a  statistical operator  
in the subspace ${{{\cal H}^0_{\scriptscriptstyle F}}}$,  
representing the  
macrosystem,  
and therefore  
        \[  
        {a_{{f}}}{{\varrho}^{\text{m}}}=0 \quad  
        {{\varrho}^{\text{m}}}  
        {a^{\scriptscriptstyle \dagger}_{f}}=0 \quad  
        \forall f  
        ,  
        \]  
while  
${\varrho}$ is a  statistical operator in the subspace  
${{{\cal H}^1_{\scriptscriptstyle F}}}$  of  
${{{\cal H}_{\scriptscriptstyle F}}}$  in which ${N}=1$.  
The coefficients  
${{\varrho}}_{gf}$ build a positive, trace one  matrix,  
which can be considered as the  
representative of a statistical operator ${\hat {\varrho}}$ in  
${{\cal H}^{(1)}}$.  
Being  
interested in the subdynamics of the microsystem  
we shall exploit the following reduction formula, valid for any  
operator  
of  
the form   
${A}  
        = \sum_{f,g}  
        {a^{\scriptscriptstyle \dagger}_{f}}  
        {A}_{fg}  
{a_{g}}  
        = \sum_{f,g}  
        {a^{\scriptscriptstyle \dagger}_{f}}  
\langle  
f  
\vert  
{\hat {\sf A}}  
\vert  
g  
\rangle  
{a_{g}}$:  
        \[  
        {\hbox{\rm Tr}}_{{\cal H}_{\scriptscriptstyle F}}  
        \left(  
        {{A}{\varrho}}  
        \right)  
         = \sum_{f,g} {A}_{fg}  
        {{\varrho}}_{gf}=  
        {\hbox{\rm Tr}}_{{{\cal H}^{(1)}}}  
        \left(  
        {{\hat {\sf A}} {\hat {\varrho}}}  
        \right)   .  
        \]  
We wish to determine the equation driving the time  
evolution of the  statistical operator on a time scale $\tau$ much  
longer than the  
typical duration of microphysical interactions for the  macrosystem, and  
therefore we shall approximate  
$        {  
        d {\varrho}_{gf}  
        \over  
        dt  
       }  
$ by:  
        \[  
        {  
        \Delta_{\tau} {\varrho}_{gf}(t)  
        \over  
        \tau  
        }  
        =  
        {1\over \tau}  
        \left[  
        {\varrho}_{gf}(t+\tau) -  
        {\varrho}_{gf}(t)  
        \right]  
        =  
        {1\over \tau}  
        \left[  
        {\hbox{\rm Tr}}_{{\cal H}_{\scriptscriptstyle F}}  
        \left(  
        {a^{\scriptscriptstyle \dagger}_{f}} {a_{g}}  
        e^{-{{  
        i  
        \over  
         \hbar  
        }}H\tau}  
        \varrho (t)  
        e^{{{  
        i  
        \over  
         \hbar  
        }}H\tau}  
        \right)  
        -  
        {{\varrho}}_{gf}(t)  
        \right]  
        .  
        \]  
To proceed further we will exploit  
the cyclicity of the trace operation, shifting the time  
evolution on the destruction and creation operators, thus  
working in Heisenberg picture.  
In this way no simplifying  
assumption is made  on   
the structure of ${{\varrho}^{\text{m}}}$.  
We now introduce   
the following superoperators, that is to say mappings  
acting  
on the algebra generated by creation and destruction  
operators:  
        \begin{equation}  
        {\cal H}^{'}={i \over \hbar} [{H},\cdot],  \quad  
        {\cal H}^{'}_0={i \over \hbar} [{H}_0 + {H}_{\text{m}},\cdot],  
        \quad  
        {\cal V}^{'}={i \over \hbar} [{V},\cdot]       .  
        \label{a7}  
        \end{equation}  
Making use of this mappings we  
evaluate ${e^{{\cal H}'\tau}}\left(  
        {{a^{\scriptscriptstyle \dagger}_{h}}{a_{k}}}  
        \right)  
$  
with the aid of the following integral representation:  
        \[  
        {{e^{{\cal H}'\tau}}{a_{k}}}  
        =  
        {\int_{-i\infty+\eta}^{+i\infty +  \eta}}{  
        dz  
        \over  
            2\pi i  
        }       \,   e^{z \tau}  
        {  
        {{  
        \left(  
        {{ z - {\cal H}'}}  
        \right)  
        }^{-1}}  
        {a_{k}}}  
        ,  
        \qquad  
        {e^{{\cal H}'\tau}}  
        \left(  
        {{a^{\scriptscriptstyle \dagger}_{h}}{a_{k}}}  
        \right)  
        =  
        \left(  
        {e^{{\cal H}'\tau}}{a^{\scriptscriptstyle \dagger}_{h}}  
        \right)  
               \left(  
        {{e^{{\cal H}'\tau}}{a_{k}}}  
               \right)  
        .  
        \]  
Let us stress at this point the relevance of the formalism  
of second quantization. The operator quantities of interest  
can be expressed in terms of products of creation and  
destruction operators. The study of their time  
evolution may thus be reconducted to evaluation of field  
operators of the form  
${e^{{\cal H}'\tau}}{a^{\scriptscriptstyle \dagger}_{h}}$  
connecting in the Fock space subspaces with $n$ and $n+1$  
particles (and similarly for  
${e^{{\cal H}'\tau}}{a_{k}}$ connecting subspaces with $n$  
and $n-1$  
particles). Thus, even recovering at the end the usual one  
particle  quantum mechanics, the Fock space structure plays  
a central role and accounts for the similarities between  
this simple case and the description of macroscopic  
systems~\cite{japan,berlin}.  
For the mappings defined in (\ref{a7}) identities hold that are  
reminiscent of the usual ones in scattering theory:  
        \begin{equation}  
        {{  
        \left(  
        {{ z - {\cal H}'}}  
        \right)  
        }^{-1}}  
        = {{  
        \left(  
        {{ z - {\cal H}'_0}}  
        \right)  
        }^{-1}}  
        \left[{1+{\cal V}'{{  
        {\left( 
        {{ z - {\cal H}'}}  
        \right)} 
        }^{-1}}}\right] =  
        {\left[{1+{{ 
        {\left( 
        {{ z - {\cal H}'}}  
        \right)} 
        }^{-1}}{\cal V}'}\right]}{{ 
        \left(  
        {{ z - {\cal H}'_0}}  
        \right)  
        }^{-1}}  .  
        \label{a9}  
        \end{equation}  
In particular we can introduce the superoperator  
${{\cal T}(z)}$  
        \begin{equation}  
        {\cal T}(z)  
        \equiv  
        {\cal V}' + {\cal V}'{{  
        \left(  
        {{ z - {\cal H}'}}  
        \right)  
        }^{-1}}{\cal V}',  
        \label{a10}  
        \end{equation}  
satisfying  
        \[  
        {{  
        \left(  
        {{ z - {\cal H}'}}  
        \right)  
        }^{-1}}={{  
        \left(  
        {{ z - {\cal H}'_0}}  
        \right)  
        }^{-1}} +{{  
        \left( 
        {{ z - {\cal H}'_0}}  
        \right) 
        }^{-1}}  
        {\cal T}(z){{  
        \left(  
        {{ z - {\cal H}'_0}}  
        \right)  
        }^{-1}}  
        \]  
and  
        \begin{equation}  
        {\cal T}(z)=  
        {\cal V}' + {\cal V}'{{  
        \left(  
        {{ z - {\cal H}'_0}}  
        \right)  
        }^{-1}}{\cal T}(z),  
        \label{a14}  
        \end{equation}  
corresponding to the Lippman-Schwinger equation for the T-matrix.  
Taking into account the fact that $[{H},{N}]=0$ one can see  
that  
the restriction to ${{{\cal H}^1_{\scriptscriptstyle F}}}$  of the  
operator ${{\cal T}(z)}{a_{k}}$ has the  
simple general form:  
        \begin{equation}  
        \label{a13}  
        i\hbar  
        {{{\cal T}(z)}{a_{k}}}_{|{{{\cal H}^1_{\scriptscriptstyle F}}}}  
        =\sum_h  
        T{}_{h}^{k}  
        \left(  
         i\hbar  z  
        \right)  
        {a_{{h}}} ,  
        \end{equation}  
where $  
T{}_{h}^{k}  
\left(  
   z  
\right)  
$ is an operator in the subspace  
${{{\cal H}^0_{\scriptscriptstyle F}}}$.  
This restriction is the only part of interest to us, since we are  
considering a single microsystem.  
Our formalism points to this matrix, whose entries are operators  
on the Hilbert space of the macrosystem, as the basic  
mathematical tool to describe the physics of the  microsystem:  
we will show that it yields all relevant quantities and, in our  
opinion, could be a sound starting point for phenomenological  
assumptions.  
$  
T{}_{h}^{k}  
\left(  
   z  
\right)  
$ bears a connection to scattering theory, as it is clear from  
(\ref{a14}); it is also related to the thermodynamics of the  
macrosystem being an operator on  
${{{\cal H}^0_{\scriptscriptstyle F}}}$.  
To help clarifying this connection we consider a simple case  
in which   
$  
T{}_{h}^{k}  
\left(  
   z  
\right)  
$  
can be explicitly calculated. Let the  macrosystem be composed of  
free particles:  
        \[  
        {H}_{\text{m}}  
         = \sum_\eta  
        {E_\eta} {b^{\scriptscriptstyle \dagger}_{\eta}}  
        {b_{{\eta}}}       ,  
        \qquad  
        V=\sum_{p,\xi, q,\eta}  
        {a^{\scriptscriptstyle \dagger}_{p}}  
        {b^{\scriptscriptstyle \dagger}_{\xi}}  
        {a^{\scriptscriptstyle }_{q}}  
        {b^{\scriptscriptstyle }_{\eta}}  
        V_{p\xi q\eta}  
        \]  
where ${b^{\scriptscriptstyle \dagger}_{\eta}}$ is the creation  
operator of a particle in an eigenstate $v_\eta$ with energy  
${E_\eta}$ (either a Bose or a Fermi particle).  
Recalling that we are describing a single particle and exploiting  
the superoperators introduced in (\ref{a7}), (\ref{a9})  
and (\ref{a10}) we can calculate  
$  
T{}_{h}^{k}  
\left(  
   z  
\right)  
$  
as defined by (\ref{a13}). To do this we bring to normal order the  
creation and destruction operators associated with the  
macrosystem  and restrict ourselves to a one mode dynamics, in  
which, apart from statistical corrections, only one creation and  
one destruction operator of the type $b$ appear: that is to say  
we neglect three particle collisions. Then one obtains:  
        \begin{eqnarray*}  
        &&  
        { T{}_{f}^{k}  
        ({E_k}+i{\varepsilon})  
        }=  
        \sum_{\xi,\eta}  
        {b^{\scriptscriptstyle \dagger}_{\xi}}  
        \langle  
        k,\xi  
        \vert  
        V^{(2)} +  
        V^{(2)}  
        {  
        1  
        \over  
        E_\xi + E_k + i\varepsilon -  
        \left(  
        H^{(2)}_0  
        +  
        V_L  
        \right)  
        }  
        V_L  
        \vert  
        f,\eta  
        \rangle  
        {b^{\scriptscriptstyle }_{\eta}}  
        ,  
        \end{eqnarray*}  
where $\varepsilon$ is a positive quantity and the  
following relationships hold,  
        \begin{eqnarray}  
        \label{n2}  
        \langle  
        k,\xi  
        \vert  
        H^{(2)}_0  
        \vert  
        f,\eta  
        \rangle  
        &=&  
        (E_f + E_\eta) \delta_{kf} \delta_{\xi\eta}  
        \nonumber  
        \\  
        \langle  
        k,\xi  
        \vert  
        V_L  
        \vert  
        f,\eta  
        \rangle  
        &=&  
        (1\pm {b^{\scriptscriptstyle \dagger}_{\xi}}  
        {b_{{\xi}}})  
        \langle  
        k,\xi  
        \vert  
        V^{(2)}  
        \vert  
        f,\eta  
        \rangle  
        =  
        (1\pm {b^{\scriptscriptstyle \dagger}_{\xi}}  
        {b_{{\xi}}})  
        V_{k\xi f\eta}  
        ;  
        \end{eqnarray}  
here the superscript $(2)$ denotes operators in the two-particle  
Hilbert space  
and statistical corrections for scattering in the  
medium  
are taken into account in the potential term  $V_L$,  
implicitly defined by (\ref{n2}) and by the usual resolvent  
series  
($+,-$ sign stand for Bose and Fermi statistics  
respectively). The connection to the familiar T-matrix is  
evident.  
\par  
We now come to the master  
equation describing the irreversible time evolution of the  
 statistical operator on the chosen time scale:  
        \begin{equation}  
        \label{Lind}  
        {  
        d {\varrho}_{kh}  
        \over  
        d\tau  
        }  
        =  
        -{i \over \hbar}  
        \left(  
        {{E_k}-{E_h}}  
        \right)  
          {\varrho}_{kh}  
        -  
        {i \over \hbar}  
        \sum_f  {{Q}}_{kf} {\varrho}_{fh}  
        +  
        {i \over \hbar}  
        \sum_g  {\varrho}_{kg} {{Q}}^{*}_{hg}  
        +  
        {1 \over \hbar}  
        \sum_{fg \atop \lambda\xi}  
        \left(  
          {{{ L}}_{\lambda\xi}}  
          \right)  
          _{kf}  
        {\varrho}_{fg}  
        {\left( 
        {{{L}}_{\lambda\xi}}  
        \right)}^{*}_{hg} 
        ,  
        \end{equation}  
from which we can read off the structure of the generator of the  
semigroup driving the time evolution.  
The quantities appearing in (\ref{Lind}) are defined in the  
following way:  
        \begin{eqnarray}  
        \label{l}  
        {{ Q}}_{kf}  
        &=&  
        {\hbox{\rm Tr}}_{{{{\cal H}_{\scriptscriptstyle F}}}}  
        \left[  
        {  
        {  
        T{}_{f}^{k}  
        ({E_k}+i{\varepsilon})  
        }  
        {{\varrho}^{\text{m}}(\tau)}  
        }  
        \right]  
        \nonumber  
        \\  
        {{Q}}^{*}_{hg}  
        &=&  
        {\hbox{\rm Tr}}_{{{{\cal H}_{\scriptscriptstyle  
        F}}}}  
        \left[  
        {  
        T{}_{g}^{h}{}^{\scriptscriptstyle \dagger}  
        (  
        {{E_h}+i{\varepsilon}}  
        )  
        {{\varrho}^{\text{m}}(\tau)}  
        }  
        \right]  
        \nonumber  
        \\  
        {\left( 
          {{{L}}_{\lambda\xi}}  
        \right)}_{kf} 
        &=&  
        \sqrt{2\varepsilon  \pi_\xi}  
        {  
        \langle  
        \lambda  
        \vert  
        {  
        T{}_{f}^{k}  
        ({E_k}+i{\varepsilon})  
        }  
        \vert  
        \xi (t)  
        \rangle  
        \over  
        {{E_k}+{E_{{\lambda}}}-{E_f}-{E}_{\xi} -i\varepsilon}  
        }  
        ,  
        \end{eqnarray}  
with $\varepsilon$ a positive  constant  
and  
$\xi(\tau)$ a complete system of  
eigenvectors of  
${{\varrho}^{\text{m}}(\tau)}$ with eigenvalues  
$ \pi_{\xi(\tau)} $.  
If we now introduce in ${{\cal H}^{(1)}}$   the operators  
${\hat {\sf H}_0},  
{\hat {\sf Q}},  
{\hat {{\sf L}}}_{\lambda\xi}$ and ${\hat \varrho}$,  
        \[  
        \langle  
        {g}  
        \vert  
        {\hat {\sf H}}_0  
        \vert  
        {f}  
        \rangle  
        =E_f \delta_{gf}  
        ,  
        \quad  
        \langle  
        {g}  
        \vert  
        {\hat {\sf Q}}  
        \vert  
        {f}  
        \rangle  
        ={Q}_{gf}  
        ,  
        \quad  
        \langle  
        {g}  
        \vert  
        {\hat {{\sf L}}_{\lambda\xi}}  
        \vert  
        {f}  
        \rangle  
        =  
         {\bigl( {{{L}}_{\lambda\xi}} \bigr)}_{gf},  
        \quad  
        \langle  
        {g}  
        \vert  
        {\hat \varrho}  
        \vert  
        {f}  
        \rangle  
        =  
        \varrho_{gf},  
        \]  
eq.\ (\ref{Lind}) becomes:  
        \[  
        {  
        d {\hat \varrho}  
        (\tau)  
        \over  
                      d\tau  
        }  
        =  
        -{i \over \hbar}  
        \left[{\hat {{\sf H}}}_0  
        +  
        {\hat {{\sf H}}}_{\text{eff}},  
        {\hat \varrho}  
        (\tau)\right]  
        -{1\over\hbar}  
        \left \{  
        {  
                {\hat \Gamma}  
                , {\hat \varrho}  
                (\tau)}  
                \right \}  
        +  
        {1 \over \hbar}  
        \sum^{}_{{\xi,\lambda  }}  
        {\hat {\sf L}}_{\lambda\xi} {\hat {\varrho}}(\tau)  
        {{\hat {\sf L}}{}_{\lambda\xi}^{\scriptscriptstyle \dagger}}\ ,  
        \]  
where  
        \[  
        {\hat {\sf H}}_{\text{eff}}=  
        {  
        {{\hat {\sf Q}}+  
        {\hat {\sf Q}}{}^{\scriptscriptstyle \dagger}}  
        \over  
        2  
        }  
        , \qquad  
        {\hat \Gamma}=i  
        {  
        {  
        {\hat {{\sf Q}}}  
        -{\hat {\sf Q}}{}^{\scriptscriptstyle \dagger}}  
        \over  
        2  
        }  
         .  
        \]  
Verification of the conservation of the trace of the  
statistical operator within the adopted approximations leads  
to the following relationship  
        \begin{equation}  
        \label{n3}  
        {\hat \Gamma}  
        \approx  
        {1\over 2}  
        \sum^{}_{{\xi,\lambda  }}  
        {{\hat {\sf L}}{}_{\lambda\xi}^{\scriptscriptstyle \dagger}}  
        {\hat {\sf L}}{}_{\lambda\xi}  
        ,  
        \end{equation}  
and therefore to:  
        \begin{equation}  
        \label{n4}  
        {  
        d {\hat \varrho}  
        (\tau)  
        \over  
                      d\tau  
        }  
        =  
        -{i \over \hbar}  
        \left[{\hat {{\sf H}}}_0  
        +  
        {\hat {{\sf H}}}_{\text{eff}},  
        {\hat \varrho}  
        (\tau)\right]  
        -{1\over\hbar}  
        \left \{  
        {  
        {1\over 2}  
        \sum^{}_{{\xi,\lambda  }}  
        {{\hat {\sf L}}{}_{\lambda\xi}^{\scriptscriptstyle \dagger}}  
        {\hat {\sf L}}{}_{\lambda\xi}  
                , {\hat \varrho}  
                (\tau)}  
                \right \}  
        +  
        {1 \over \hbar}  
        \sum^{}_{{\xi,\lambda  }}  
        {\hat {\sf L}}_{\lambda\xi} {\hat {\varrho}}(\tau)  
        {{\hat {\sf L}}{}_{\lambda\xi}^{\scriptscriptstyle  
        \dagger}}\ .  
        \end{equation}  
This master-equation is a typical result of the formalism  
restricted to the case of a single microsystem, for the general structure  
see~\cite{japan,berlin} 
\par  
Before applying  
(\ref{n4}) to a concrete physical  
situation  
it can be useful to gain some further insight into the structure  
of the  operators appearing in it. As already said the quantity  
that the formalism suggests as a natural candidate where to put  
in suitable phenomenological expressions is the  operator  
$ T{}_{f}^{k} \left( z \right)$, an  operator whose trace over the 
Fock space for the  macrosystem calculated with  
${\varrho}^{\text{m}}$ gives the value of the T-matrix for  
scattering  
from state $u_f$ to state $u_k$ averaged over the state of the  
macroscopic system. A quite general phenomenological expression  
may  be obtained in the following way.  
Suppose that         ${{\cal T}(z)}$ has the form  
        \[  
        {{\cal T}(z)}  
        =  
        {i\over \hbar}  
        \left[  
        V(i\hbar z),\cdot  
        \right]  
                  ,  
        \qquad  
        V(z)=  
        \sum_{k\lambda f \mu}  
        V_{k\lambda f \mu}(z)  
        a^{\scriptscriptstyle\dagger}_{k}  
        b^{\scriptscriptstyle\dagger}_{\lambda}  
        a_f  
        b_\mu  
        ,  
        \]  
with $b^{\scriptscriptstyle\dagger}$, $b$ creation and  
destruction  operators in the Fock space for the  macrosystem.  
We thus have  
        \[  
        i\hbar  
        {{{\cal T}(z)}{a_{k}}}  
        =  
        \sum_{\lambda f \mu}  
        V_{k \lambda f \mu}(i\hbar z)  
        b^{\scriptscriptstyle\dagger}_{\lambda}  
        a_f  
        b_\mu  
        =  
        \sum_{f}  
        T{}_{f}^{k} \left( i\hbar z \right) 
        a_f  
        ,  
        \]  
and supposing translation invariance in the interaction kernel:  
        \begin{equation}  
        \label{tpheno}  
        T{}_{f}^{k} \left( z \right) 
        =  
        \sum_{\lambda  \mu}  
        b^{\scriptscriptstyle\dagger}_{\lambda}  
        V_{k\lambda f \mu}(z)  
        b_\mu  
        =  
        {\int d^3 \! {\bbox{x}} \,}  
        {\int d^3 \! {\bbox{y}} \,}  
        \psi^{\scriptscriptstyle\dagger}({\bbox{x}})  
        u_k^{*}({\bbox{y}})  
        t(z,{\bbox{x}}-{\bbox{y}})  
        u_f({\bbox{y}})  
        \psi({\bbox{x}})  
        .  
        \end{equation}  
Such an Ansatz amounts to introduce an effective potential  
which should give in Born approximation the full scattering  
amplitude.  
As a result the potential term in (\ref{n4}) is linked to the  
scattering amplitude, as we shall see in the next paragraph,  
while the incoherent contribution is generally connected to the  
scattering cross section.  
To realize this let us consider the last term of (\ref{Lind}),  
keeping the proposed Ansatz into account:  
        \begin{eqnarray}  
        \label{guaio}  
        &&  
        {  
        2\varepsilon  
        \over  
                    \hbar  
        }  
        \sum^{}_{{\lambda,\lambda'\atop \lambda''}}  
        \sum_{f,g}  
        {\int d^3 \! {\bbox{x}} }  
        {\int d^3 \! {\bbox{y}} \,}  
        u_k^{*}({\bbox{y}})  
        {  
        t  
        {
	\left(  
        E_k +i\varepsilon,{\bbox{x}}-{\bbox{y}}  
        \right)  
        }
	\over  
        {{E_k}+{E_{{\lambda}''}}-{E_f}-{E_{{\lambda}}}    -i\varepsilon}  
        }  
        u_f({\bbox{y}})  
        \langle  
        {\lambda''}  
        \vert  
        {\psi}^{\scriptscriptstyle\dagger}({\bbox{x}})  
        {\psi}({\bbox{x}})  
        \vert  
        {\lambda}  
        \rangle  
        \langle  
        \lambda  
        {
	\left | 
        {\varrho}^{\text{m}}(\tau)  
        \right |
	} 
        {\lambda}'  
        \rangle  
        \nonumber \\  
        &&  
        \times  
        \varrho_{fg}(\tau)  
        {\int d^3 \! {\bbox{x}'} }  
        {\int d^3 \! {\bbox{y}'} \,}  
        \langle  
        {\lambda'}  
        \vert  
        {\psi}^{\scriptscriptstyle\dagger}({\bbox{x}'})  
        {\psi}({\bbox{x}'})  
        \vert  
        {\lambda''}  
        \rangle  
        u_g^{*}({\bbox{y}'})  
        {  
        t^{*}  
        {
	\left(  
        E_h +i\varepsilon,{\bbox{x}'}-{\bbox{y}'}  
        \right)
	}  
        \over  
        {{E_h}+{E_{{\lambda}''}}-{E_g}-{E_{{\lambda}'}}    +i\varepsilon}  
        }  
        u_h({\bbox{y}'})  
        ,  
        \end{eqnarray}  
and let us specialize to the case of a diagonal matrix element.  
Supposing the  statistical operator for the  microsystem is  
quasi-diagonal and the  macrosystem is at equilibrium, so that  
$  
\varrho^{\rm m}|\lambda\rangle  
=\varrho^{\rm m}_\lambda |\lambda\rangle  
$,  
we exploit the usual representation for the delta function, thus  
obtaining:  
        \begin{eqnarray}  
        &&  
        \sum_{f}  
        \sum_{\lambda \lambda'}  
        {  
        2\pi  
        \over  
            \hbar  
        }  
        \delta  
        {
	\left(  
        E_k + E_\lambda -E_f -E_{\lambda'}  
        \right)  
        }
	\nonumber  
        \\  
        &&  
        \hphantom{  
        \sum_{\lambda}  
        \sum_{f \lambda'}  
        \times  
        }  
        \times  
        {
	\left | 
        {\int d^3 \! {\bbox{x}} \,}  
        {\int d^3 \! {\bbox{y}} \,}  
        u_k^{*}({\bbox{y}})  
        \langle  
        {\lambda}  
        \vert  
        {\psi}^{\scriptscriptstyle\dagger}({\bbox{x}})  
        t  
        {
	\left(  
        E_k +i\varepsilon,{\bbox{x}}-{\bbox{y}}  
        \right)  
        }
	{\psi}({\bbox{x}})  
        \vert  
        {\lambda'}  
        \rangle  
        u_f({\bbox{y}})  
        \right |
	}^2 
        \varrho^{\rm m}_{\lambda'}  
        \varrho_{ff}(\tau)  
        .  
        \nonumber  
        \end{eqnarray}  
In this formula one has the typical transition probability  
between an initial state  
$f,\lambda'$  
and a final state  
$k,\lambda$, averaged over all possible initial configurations  
and summed over all possible final  
states for the  macrosystem, that is to say  contributions  
from both  coherent and diffuse scattering are included.  
It might  be instructive  to show in a different way the  
connection between the last term of (\ref{Lind}) and the  
total scattering cross section, referring to a famous paper  
by van Hove~\cite{vanHove}. Taking for concreteness the  
Fermi  pseudopotential (see next paragraph), whose Fourier  
transform is simply the constant  
${\tilde V}={  
2\pi\hbar^2  
\over  
           m  
}b$, we evaluate the diagonal element of  (\ref{guaio})  
assuming that the $u_f$ are given by plane waves (the  
indexes $f,g,h,k$ becoming momenta), thus obtaining  
[$N({\bbox{x}})=  
        {\psi}^{\scriptscriptstyle\dagger}({\bbox{x}})  
        {\psi}({\bbox{x}})  
$]:  
        \begin{eqnarray*}  
        &&  
        {  
        2\varepsilon  
        \over  
                    \hbar  
        }  
        {\left | 
        {\widetilde V}  
        \right |}^2 
        \sum^{}_{{\lambda,\lambda'}}  
        {\int  
        {  
        d^3 \! {\bbox{P}}  
        \over  
        (2\pi\hbar)^3  
        }  
        }  
        {\int  
        {  
        d^3 \! {\bbox{q}}  
        \over  
        (2\pi\hbar)^3  
        }  
        }  
        {\int d^3 \! {\bbox{x}} }  
        {\int d^3 \! {\bbox{y}} \,}  
        \\  
        &&  
        \hphantom{        {  
        2\varepsilon  
        \over  
                    \hbar  
        }  
        {\left | 
        {\widetilde V}  
        \right |}^2 
        \sum^{}_{{\lambda,\lambda'}}  
        }  
        \times  
        e^{  
        - {i\over\hbar}  
        \left[  
                      {\bbox{k}} -  
                      \left(  
                      {\bbox{P}}+{  
                      {\bbox{q}}  
                      \over  
                              2  
                      }  
                      \right)  
        \right]  
        \cdot  
        {\bbox{x}}  
        }  
        {  
        \langle  
        {\lambda'}  
        \vert    N({\bbox{x}})  
        \vert  
        {\lambda}  
        \rangle  
        \over  
        E_k + E_{\lambda'}-{  
        1  
        \over  
         2m  
        }  
        {\left( 
        {\bbox{P}}+\frac 12 {\bbox{q}}  
        \right)}^2   - E_{\lambda} -i\varepsilon 
        }  
        \langle  
        {\bbox{P}}+\frac 12 {\bbox{q}}  
        |  
        {\hat {\varrho}}  
        |  
        {\bbox{P}}-\frac 12 {\bbox{q}}  
        \rangle  
        \\  
        &&  
        \hphantom{        {  
        2\varepsilon  
        \over  
                    \hbar  
        }  
        {\left | 
        {\widetilde V}  
        \right |}^2 
        \sum^{}_{{\lambda,\lambda'}}  
        }  
        \times  
        \langle  
        \lambda  
        \left |  
        {\varrho}^{\text{m}}(\tau)  
        \right |  
        {\lambda}  
        \rangle  
        {  
         \langle  
        {\lambda}  
        \vert  
        N({\bbox{y}})  
        \vert  
        {\lambda'}  
        \rangle  
        \over  
        E_k + E_{\lambda'}-{  
        1  
        \over  
         2m  
        }  
        {\left( 
        {\bbox{P}}-\frac 12 {\bbox{q}}  
        \right)}^2   - E_{\lambda} +i\varepsilon 
        }  
        e^{  
         {i\over\hbar}  
        \left[  
                      {\bbox{k}} -  
                      \left(  
                      {\bbox{P}}-{\bbox{q} \over 2}  
                      \right)  
        \right]  
        \cdot  
        {\bbox{y}}  
        }        ,  
        \end{eqnarray*}  
and supposing ${\hat {\varrho}}$  such that the energies in the  
denominators may be considered approximately equal,  
introducing the Wigner function for the neutron  
        \[  
        f_{\text{w}} ({\bbox{x}},{\bbox{p}})  
        =  
        {\int  
        {  
        d^3 \! {\bbox{q}}  
        \over  
        (2\pi\hbar)^3  
        }  
        }  
        e^{  
         {i\over\hbar}  
                      {\bbox{x}}  
        \cdot  
        {\bbox{q}}  
        }  
        \langle  
        {\bbox{p}}+\frac 12 {\bbox{q}}  
        |  
        {\hat {\varrho}}  
        |  
        {\bbox{p}}-\frac 12 {\bbox{q}}  
        \rangle  
        \]  
one easily has  
        \begin{eqnarray}  
        \label{wigner}  
        &&  
        {  
        2\pi  
        \over  
            \hbar  
        }  
        {\left( 
        {  
        1  
        \over  
         2\pi\hbar  
        }  
        \right)}^4 
        {\left | 
        {\widetilde V}  
        \right |}^2 
        {\int d^3 \! {\bbox{P}} }  
        {\int dt}  
        {\int d^3 \! {\bbox{r}} }  
        \,  
        e^{  
        -{  
        i  
        \over  
         \hbar  
        }  
        \left(  
        {  
        P^2  
        \over  
           2m  
        }    - E_k  
        \right)   t  
        +  
        {  
        i  
        \over  
         \hbar  
        }  
        ({\bbox{P}}-{\bbox{k}}) \cdot {\bbox{r}}  
        }  
        \nonumber \\  
        &&  
        \hphantom{        {  
        2\pi  
        \over  
            \hbar  
        }  
        {\left( 
        {  
        1  
        \over  
         2\pi\hbar  
        }  
        \right)}^4 
        {\left | 
        {\widetilde V}  
        \right |}^2 
        {\int d^3 \! {\bbox{P}} }  
        {\int dt}  
        {\int d^3 \! {\bbox{r}} }  
        }  
        \times  
        {\int d^3 \! {\bbox{X}} }  
        f_{\text{w}}({\bbox{X}},{\bbox{P}})  
        \langle  
        N  
        \left(  
        {\bbox{X}} -  
        {  
        {\bbox{r}}  
        \over  
                2  
        }  
        \right)  
        N  
        \left(  
        {\bbox{X}} +  
        {  
        {\bbox{r}}  
        \over  
                2  
        }    ,t  
        \right)  
        \rangle    ,  
        \end{eqnarray}  
where $  
\left \langle  
\ldots  
\right \rangle \equiv  
{\mbox{\rm Tr}}_{{{{\cal H}_{\scriptscriptstyle F}}}}  
\left(  
\ldots \varrho^{\text{m}}  
\right)  
$, and  
$  N({\bbox{x}},t)  
$                denotes the  operator in the Heisenberg  
picture.  
We have thus recovered the typical factorized structure  
appearing in the expression for the scattering cross section  
of a neutron off a macroscopic system: square modulus of the  
Fourier transform of the interaction potential times the  
dynamic structure function depending on transferred momentum  
and energy, with the refinement that it is here  
weighted according to position and momentum distribution of  
the incoming particle.  
For the non-diagonal matrix element one  
can expect to obtain analogous results if the quantities  
appearing in (\ref{guaio}) are sufficiently slowly varying functions of  
their arguments, so that, in the continuous limit, an  
interpolation formula of the form  
        \[  
        \varepsilon  
        \int d\xi \,  
        {  
        g(\xi)  
        \over  
        \left(  
        \alpha+\xi+i\varepsilon  
        \right)  
        \left(  
        \beta+\xi-i\varepsilon  
        \right)  
        }  
        \approx \pi \int d\xi \, g(\xi)\delta(\beta+\xi)  
        \approx \pi \int d\xi \, g(\xi)\delta(\alpha+\xi)  
        ,  
        \qquad  
        \left |  
        \alpha-\beta  
        \right |    \ll \varepsilon  
        \]  
with $g(\xi)$ a suitably smooth function may be used. The failure  
of such an approximation and thus the relevance of the actual  
value of the parameter $\varepsilon$ in the final expression  
might be traced back to the breakdown of the approximations that  
have led to the markovian evolution generated by the  
master-equation (\ref{Lind}).  
\section{OPTICAL BEHAVIOR}  
We now devote our attention to the interaction of neutrons with  
matter. This field is well suited to test our formalism both  
because of the very refined experiments that have been carried  
out in neutron interferometry~\cite{Erice,Rauch} and because of the very  
well studied description of neutron optics phenomena, as  
developed for example in the book by Sears~\cite{Sears},  
that we  
will take as basic reference. As a first step we want to consider  
the  coherent interaction of neutrons with matter and therefore we  
neglect in (\ref{Lind}) the last contribution, linked to  
incoherent processes. As we will see later this term implies  
indeed a smaller correction in the case of neutron scattering.  
We are left with:  
        \begin{equation}  
        \label{b1}  
        {  
        d {\varrho}_{kh}  
        \over  
        d\tau  
        }  
        =  
        -{i \over \hbar}  
        \left(  
        {{E_k}-{E_h}}  
        \right)  
          {\varrho}_{kh}  
        -  
        {i \over \hbar}  
        \sum_f  {{Q}}_{kf} {\varrho}_{fh}  
        +  
        {i \over \hbar}  
        \sum_g  {\varrho}_{kg} {{Q}}^{*}_{hg}  
        ,  
        \end{equation}  
and we need a suitable expression for the operator  
        \[  
        {{Q}}_{kf}  
        =  
        {\hbox{\rm Tr}}_{{{{\cal H}_{\scriptscriptstyle F}}}}  
        \left[  
        {  
        {  
        T{}_{f}^{k}  
        ({E_k}+i{\varepsilon})  
        }  
        {{\varrho}^{\text{m}}(\tau)}  
        }  
        \right]  
        .  
        \]  
Following Sears we adopt the Fermi  
pseudopotential to describe the neutron nucleus interaction in  
impulse approximation; let us recall the form of the  
T-matrix in the context of the  
elementary theory of dispersion:  
        \begin{equation}  
        T={  
        2\pi \hbar^2  
        \over  
        m  
        }  
        \sum_\alpha b_\alpha  
        \sum_{i=1}^{N_\alpha}  
         \delta^3 ({\hat {{\sf x}}}-{\bbox{R}_i}),  
        \label{b4}  
        \end{equation}  
where ${\hat {{\sf x}}}$ is the position operator for the  
neutron, ${\bbox{R}_i}$ the position operator for the i-th nucleus of type  
$\alpha$, $b_\alpha$ the  
bound scattering length, depending on isotope and spin  
orientation, $m$ the neutron mass, $N_\alpha$ the number of  
nuclei of type $\alpha$.  
An operator of the form (\ref{b4}), that is to say a sum over  
one-particle operators, is expressed in  
second quantization by:  
        \begin{equation}  
        \label{potsears}  
        T=  
        {  
        2\pi \hbar^2  
        \over  
        m  
        }  
        \sum_\alpha b_\alpha  
        {\int d^3 \! {\bbox{x}} \,}  
        {\psi}^{\scriptscriptstyle\dagger}_\alpha ({\bbox{x}})  
        \delta^3 ({\hat {{\sf x}}}-{{\bbox{x}}})  
        {\psi}_\alpha ({\bbox{x}})  
        ,  
        \end{equation}  
where ${\psi}_\alpha ({\bbox{x}})$ is the  
field operator, acting in the Fock space of the  macrosystem,  
corresponding to particles of type $\alpha$.  
For the sake of simplicity from now on  we will consider  
one kind of particles, thus dropping the subscript $\alpha$.  
Furthermore we  
will assume that $b$ is a real quantity, since we are not going to  
deal with absorption phenomena. As we shall see in the next  
section we concentrate on non-hermiticity of the potential  
connected with  incoherent processes and not with net absorption.  
A phenomenological  description as given by (\ref{potsears})  
falls within the class of effective potentials considered  
in the previous paragraph and corresponds to the following  
interaction kernel:  
        \begin{equation}  
        \label{fermi}  
        t(z,{\bbox{x}}-{\bbox{y}})=  
        {  
        2\pi \hbar^2  
        \over  
        m  
        }     b  
        \,  
        \delta^3 ({\bbox{x}}-{{\bbox{y}}})  
        ,  
        \end{equation}  
leading to  
        \[  
        {  
        T{}_{f}^{k}  
        ({E_k}+i{\varepsilon})  
        }  
        =  
        {  
        2\pi \hbar^2  
        \over  
        m  
        }     b  
        {\int d^3 \! {\bbox{x}} \,}  
        {\int d^3 \! {\bbox{y}} \,}  
        {\psi}^{\scriptscriptstyle\dagger}({\bbox{x}})  
        u_k^{*}({\bbox{y}})  
        \delta^3 ({\bbox{x}}-{{\bbox{y}}})  
        u_f({\bbox{y}})  
        {\psi}({\bbox{x}})  
                        .  
        \]  
Eq.\ (\ref{b1}) thus becomes, in operator  
form:  
        \begin{eqnarray}  
        \label{b8}  
        {  
        d {\hat {\varrho}}(\tau)  
        \over  
                      dt  
        }           =  
        &-&  
        {i \over \hbar}  
        \left[{\hat {{\sf H}}_0,{\hat {\varrho}}(\tau)}\right]  
        -{  
        i  
        \over  
         \hbar  
        }  
        {  
        2\pi \hbar^2  
        \over  
        m  
        }     b  
        {\int d^3 \! {\bbox{x}} \,}  
        {\langle  
        {\psi}^{\scriptscriptstyle\dagger}({\bbox{x}})  
        {\psi}({\bbox{x}})  
        \rangle}_{\tau} \,  
        \delta^3 ({\hat {{\sf x}}}-{\bbox{x}})  
        {\hat \varrho}  
        (\tau)  
        \nonumber \\  
        {}  
        &+&  
        {  
        i  
        \over  
         \hbar  
        }  
        {  
        2\pi \hbar^2  
        \over  
        m  
        }     b  
        {\int d^3 \! {\bbox{x}} \,}  
        {\langle  
        {\psi}^{\scriptscriptstyle\dagger}({\bbox{x}})  
        {\psi}({\bbox{x}})  
        \rangle}_{\tau}   \,  
        {\hat \varrho}(\tau)  
        \delta^3 ({\hat {{\sf x}}}-{\bbox{x}})  
        ,  
        \end{eqnarray}  
where ${\hat {{\sf x}}}$ is the position operator for the neutron and  
${  
\langle  
A  
\rangle  
}_\tau  \equiv  
        {\hbox{\rm Tr}}_{{{{\cal H}_{\scriptscriptstyle F}}}}  
({{ {\varrho}}^{\text{m}}(\tau)}  
A)  
$. If we consider only pure states  
and assume the  
macrosystem to be at equilibrium  
(  
$  
{\langle  
\ldots  
\rangle  
}_\tau \equiv \langle  
\ldots  
\rangle  
$  
), eq.\ (\ref{b8}) is equivalent to  
the following stationary Schr\"odinger equation  
        \begin{equation}  
        \label{Sears}  
        \left \{  
        -  
        {  
        \hbar^2  
        \over  
               2m  
        }  
        \Delta_x  
        +  
        {  
        2\pi \hbar^2  
        \over  
        m  
        }     b  
        {\langle  
        {\psi}^{\scriptscriptstyle\dagger}({\bbox{x}})  
        {\psi}({\bbox{x}})  
        \rangle}  
        \right \}  
        \phi({\bbox{x}})  
        = E  
        \phi({\bbox{x}})  
        ,  
        \end{equation}  
which, remembering that the average particle density  
$  
\langle  
\sum_i \delta^3 ({\bbox{x}}-{\bbox{R}_i})  
\rangle  
$  
is given in second quantization by  
${\langle  
        {\psi}^{\scriptscriptstyle\dagger}({\bbox{x}})  
        {\psi}({\bbox{x}})  
        \rangle}$,  
is exactly the equation used by Sears  
to describe all  coherent neutron optical phenomena, here  
recovered in a straightforward, alternative way, though in a  
very different framework.  
The  
term  
        \[  
        {  
        2\pi \hbar^2  
        \over  
        m  
        }     b  
        {\langle  
        {\psi}^{\scriptscriptstyle\dagger}({\bbox{x}})  
        {\psi}({\bbox{x}})  
        \rangle}  
        \]  
is called optical potential and assumes different  
expressions according to the structure of the system. If the  
medium can be considered homogeneous, with density $n_{\text{o}}$,  
eq.\ (\ref{Sears}) describes propagation of matter waves with an  
index of refraction given by  
        \begin{equation}  
        \label{Gold}  
        n={  
        \left(  
        1-  
        {  
        2\pi \hbar^2  
        \over  
        m E  
        }     b  
        n_{\text{o}}  
        \right)  
        }^{\frac 12}  
        \simeq  
        1-  
        {  
        \lambda^2  
        \over  
                 2\pi  
        }  
        b n_{\text{o}}  
        \quad,  
        \end{equation}  
as first obtained by Goldberger and Seitz~\cite{Gold} in the  
absence of absorption. This is the formula currently used to  
calculate phase shifts in neutron interferometry  
experiments~\cite{Erice}:  
        \begin{equation}  
        \label{phaseshift}  
        e^{i \chi}  
        =  
        e^{i (n-1) {2\pi \over \lambda}D}  
        =  
        e^{-i n_{\text{o}} b \lambda D}  ,  
        \end{equation}  
where $D$ is the thickness of the sample.  
\par  
In a similar way we can obtain from (\ref{b1}) a more general  
formula for the refractive index introduced for the first time by  
Lax~\cite{Lax}.  
Starting from the general expression (\ref{tpheno}) the potential  
term in (\ref{b1}) becomes  
        \[  
        \sum_f  {{Q}}_{kf} {\varrho}_{fh}(\tau)  
        =  
        \sum_f  
        {\hbox{\rm Tr}}_{{{{\cal H}_{\scriptscriptstyle F}}}}  
        {\int d^3 \! {\bbox{x}} \,}  
        {\int d^3 \! {\bbox{y}} \,}  
        {\psi}^{\scriptscriptstyle\dagger}({\bbox{x}})  
        u_k^{*}({\bbox{y}})  
        t(E_k+i\varepsilon,{\bbox{x}}-{\bbox{y}})  
        u_f({\bbox{y}})  
        {\psi}({\bbox{x}})  
        \varrho^{\rm m}(\tau)  
        \varrho_{fh}(\tau)  
                          .  
        \]  
Following Lax we suppose that the system is homogeneous, so  
that  
        \[  
        {\hbox{\rm Tr}}_{{{{\cal H}_{\scriptscriptstyle F}}}}  
        \left[  
        {\psi}^{\scriptscriptstyle\dagger}({\bbox{x}})  
        {\psi}({\bbox{x}})  
        \varrho^{\rm m}(\tau)  
        \right]  
        =  
        n_{\text{o}}  
        .  
        \]  
We have  
        \begin{eqnarray*}  
        \sum_f  {{Q}}_{kf} {\varrho}_{fh}(\tau)  
        &=&  
        n_{\text{o}}  
        \sum_f  
        {\int d^3 \! {\bbox{x}} \,}  
        t(E_k+i\varepsilon,{\bbox{x}})  
        {\int d^3 \! {\bbox{y}} \,}  
        u_k^{*}({\bbox{y}})  
        u_f({\bbox{y}})  
        \varrho_{fh}(\tau)  
        \\  
        &=&  
        n_{\text{o}}  
        {\int d^3 \! {\bbox{x}} \,}  
        t(E_k+i\varepsilon,{\bbox{x}})  
        \varrho_{kh}(\tau)  
                  ,  
        \end{eqnarray*}  
where we have exploited the orthogonality between the states  
$  
\left \{  
u_f  
\right \}  
$, thus obtaining the matrix element of the T-operator for  
forward scattering, averaged over the possible states of the  
macrosystem.  
Keeping the relation between T-operator and scattering  
amplitude into account we come to  
        \[  
        -  
        n_{\text{o}}  
        {  
        2\pi \hbar^2  
        \over  
        m  
        }  
        f(0,E_k)  
        {\varrho}_{kh}(\tau).  
        \]  
Inserted in the Schr\"odinger equation this term is equivalent  
to an index of refraction of the form:  
        \begin{equation}  
        \label{Lax}  
        n={  
        \left(  
        1  
        +  
        {  
        2\pi \hbar^2  
        \over  
        m E_k  
        }  
        n_{\text{o}}  
        f(0,E_k)  
        \right)  
        }^{\frac 12}  
        \simeq  
        1  
        +  
        {  
        \lambda^2  
        \over  
                 2\pi  
        }  
        n_{\text{o}}  
        f(0,E_k)  
        ,  
        \end{equation}  
simply linked to the forward scattering amplitude. An analogous  
result holds for electromagnetic waves propagating in a material  
with low density~\cite{Jackson}. A similar treatment has been  
proposed~\cite{Vigue} and adopted (see for  
example~\cite{Schmiedmayer})  
in the description of the propagation of atoms through a dilute  
medium, showing the interest of similar descriptions also for  
atom optics. In the case of thermal neutrons the scattering amplitude is  
isotropic within a very good approximation and is given in terms  
of the scattering length by the simple formula $f=-b$ which  
reduces (\ref{Lax}) to (\ref{Gold}).  
\par  
So far we have shown how, starting from (\ref{Lind}) and neglecting  
the  incoherent term,     we can recover some important results  
obtained within the framework of multiple scattering theory and  
used to describe the  coherent interaction of neutrons with  
matter. Our formalism  
puts in evidence  
the  statistical operator of the  macrosystem, the T-matrix and the  
scattering amplitude, so that  
phenomenological inputs are rather direct.  
Further improvements of the formulas obtained  
are allowed by the  
presence of ${\varrho}^{\text{m}}(\tau)$ and depend on its  
evaluation. The correction factor $c$ that Lax includes in  
(\ref{Lax}) to obtain the index of refraction  
        \[  
        n  
        \simeq  
        1  
        +  
        {  
        \lambda^2  
        \over  
                 2\pi  
        }  
        n_{\text{o}}  
        c  
        f(0,E_k)  
        ,  
        \]  
is connected to fulfilment of the optical theorem, which in our  
formalism, as we will see in the next section, is related to the  
presence of the  incoherent  contribution.  
\section{THE  INCOHERENT  CONTRIBUTION}  
We come now to the main statement of this paper,  
the connection between  
the  contributions other than the commutator in  (\ref{n4})  
and the dynamic structure function,  
together with  
the relevance of this relationship to the optical theorem. As  
observed by Sears an expression of the form (\ref{Gold}) or  
(\ref{Lax}) for the refractive index {\em doesn't include the  
contribution to the attenuation of the  coherent wave in the  
medium due to diffuse scattering and, hence, violates the  
``optical theorem'' of scattering theory}~\cite{21,Sears,Searsphysica}. To  
overcome this difficulty he refrains from ad hoc assumptions as  
in~\cite{Halpern}, which amount to introduce a suitable  
imaginary  contribution to the potential, and considers a rigorous  
theory of dispersion. In this more accurate treatment (\ref{b4})  
is replaced by  
        \[  
        T={  
        2\pi \hbar^2  
        \over  
        m  
        }  
        \sum_\alpha f_\alpha  
        \sum_{i=1}^{N_\alpha}  
         \delta^3 ({\hat {{\sf x}}}-{\bbox{R}_i}),  
        \]  
and $f_\alpha$ has the general expression (${\bbox{k}}$ is the  
incident neutron momentum)  
        \[  
        f_\alpha=-b_\alpha +  
        {  
        i  
        \over  
         \hbar  
        }  
        k b_{\alpha}^2  + O(k^2),  
        \]  
where the second term had been previously omitted because of its  
smallness, since   typically  
${  
1  
\over  
 \hbar  
}kb \leq {10}^{-4}$. Furthermore the scattering amplitude is to  
be multiplied by a constant $c$  which should  
take local field corrections into account and whose value depends  
only on the temperature, density and chemical composition of the  
medium.  
Sears obtains an estimate for this constant in terms of the  
structure function of the macroscopic scatterer in the case  
of an homogeneous medium, applying a multiple wave formalism  
to solve the scattering problem, and drawing strong  
analogies to the usual descriptions of propagation of  
electromagnetic waves.  
In this way he recovers a correspondence between  
attenuation of the  coherent wave in the medium and diffuse  
scattering. In the following we shall set $f_\alpha=f \  \forall  
\alpha$ and consider only real $b$, in order to concentrate  
upon diffuse scattering, neglecting absorption. By  
diffuse scattering we intend all scattering that is not  coherent  
in the absolute sense, that is elastic and  coherent  
(for the distinction between  
absolute and relative  incoherence see for  
example~\cite{Sears,Lax}).  
To compare with these more refined  
results we have to consider all  contributions in  
(\ref{n4}). Let us  stress from the very beginning some  
general features of this expression, thanks to which it can  
describe more general physical situations than those arising  
in an evolution driven by a Schr\"odinger-like equation. The  
last two terms  
        \begin{equation}  
        \label{aa1}  
        {}-{  
        1  
        \over  
          \hbar  
        }  
        \left \{  
        \frac 12  
        \sum^{}_{{\xi,\lambda  }}  
        {\hat {{\sf L}}}{}_{\lambda\xi}^{\scriptscriptstyle \dagger}  
        {\hat {{\sf L}}}_{\lambda\xi} , {\hat \varrho}  
        \right \}  
        +  
        {  
        1  
        \over  
         \hbar  
        }  
        \sum^{}_{{\xi,\lambda  }}  
        {\hat {{\sf L}}}_{\lambda\xi}  {\hat \varrho}  
        {\hat {{\sf L}}{}_{\lambda\xi}^{\scriptscriptstyle \dagger}}  
        \end{equation}  
allow for the presence of a non-self-adjoint  
potential which is nevertheless not linked to real  
absorption. This is the case for the present treatment, in  
which the imaginary part of the optical potential is to be  
traced back to the existence of diffuse scattering, as  
opposed to the  coherent wavelike behavior. Attenuation of  
the ``coherent wave'' is due to the presence of the  
anticommutator term,  
responsible for the imaginary potential,  
balanced by the last  contribution,  
typically incoherent in that it leads from a pure state to a  
mixture.  
This last term is given by a sum over subcollections,  
formally similar to the expression that we would obtain for  
the statistical operator after the measurement of a given  
observable (see~\cite{art1}). The subcollections are denoted  
by the indexes $\lambda\xi$, which specify a change of the  
state of the macroscopic system, caused by interaction with  
the microsystem, thus making this contribution to the  
dynamics incoherent. In fact we will see in the case of  
neutron-matter interaction that the trace of this term gives  
all the contributions to incoherent scattering, that is to  
say the total diffusion cross section.  
The balance between the two terms of  (\ref{aa1})  
accounts for fulfilment of the  
optical theorem.  
\par  
To see this let us now consider  (\ref{aa1}) in more detail.  
Starting from (\ref{l}) and (\ref{tpheno}), introducing a Laplace  
transform for the energy dependence of the effective  
T-matrix  
        \[  
        t(E,{\bbox{x}})=  
        \int_0^\infty d\sigma \,  
        e^{  
        {  
        i  
        \over  
         \hbar  
        }     E\sigma  
        }  
        \bar{t}(\sigma,{\bbox{x}}),  
        \]  
together with the following expression for the density  
number operator in terms of creation and destruction  
operators with specified momentum  
        \[  
        N({\bbox{x}}) =  
        {\psi}^{\scriptscriptstyle\dagger}({\bbox{x}})  
        {\psi}({\bbox{x}})  
        =  
        {  
        1  
        \over  
        {{\sf V}}  
        }  
        \sum_{\kappa, P}  
        e^{-{i\over \hbar}{\bbox{\kappa}}\cdot{\bbox{x}}}  
        b^{\scriptscriptstyle\dagger}_{P+{\kappa\over 2}}  
        b_{P-{\kappa\over 2}} ,  
        \]  
we obtain  
        \begin{eqnarray*}  
        {\hat {{\sf L}}}_{\lambda\xi}  
        &=&  
        {i\over\hbar}\sqrt{2\varepsilon\pi_\xi}  
        {1\over{{\sf V}}}  
        \sum_{\kappa, P}  
        \int_0^\infty d\tau \, e^{-{\varepsilon\over\hbar}\tau}  
        \int_0^\infty d\sigma \,  
        {\int d^3 \! {\bbox{x}''} \,}  
        e^{-{\varepsilon\over\hbar}\sigma}  
        e^{-{i\over\hbar}{\hat {{\sf H}}}_0 (\tau-\sigma)}  
        \bar{t}(\sigma,{{\bbox{x}}''}-{\hat {{\sf x}}})  
        e^{{i\over\hbar}{\hat {{\sf H}}}_0 \tau}  
        e^{-{i\over \hbar}{\bbox{\kappa}}\cdot{\bbox{x}''}}  
        \\  
        &&  
        \times  
        \langle  
        \lambda  
        |  
        e^{-{i\over\hbar} H_{\text{m}}\tau}  
        b^{\scriptscriptstyle\dagger}_{P+{\kappa\over 2}}  
        b_{P-{\kappa\over 2}}   
        e^{{i\over\hbar} H_{\text{m}}\tau}  
        |  
        \xi  
        \rangle  
        ,  
        \end{eqnarray*}  
where ${{\sf V}}$ is the volume of the region in which the system is  
supposed to be confined.  
Indicating by  ${\tilde t}(E,{\bbox{\kappa}})$  
the Fourier transform of the potential with respect to space  
        \[  
        {\tilde t}(E,{\bbox{\kappa}})  
        =  
        {\int d^3 \! {\bbox{x}} \,}  
        t(E,{\bbox{x}})  
        e^{-{i\over\hbar}{\bbox{\kappa}}\cdot{\bbox{x}}}  
        \]  
and after some simple manipulations one comes to  
        \begin{eqnarray*}  
        {\hat {{\sf L}}}_{\lambda\xi}  
        &=&  
        {i\over\hbar}\sqrt{2\varepsilon\pi_\xi}  
        {1\over{{\sf V}}}  
        \sum_{\kappa, P}  
        \int_0^\infty  
        \!  
        d\tau \, e^{-{\varepsilon\over\hbar}\tau}  
        \,  
        {\tilde t}  
        ({\hat {{\sf H}}}_0+i\varepsilon,{\bbox{\kappa}})  
        e^{{i\over\hbar} {{\bbox{\kappa}}^2\over 2m} \tau}  
        e^{{i\over\hbar} {{\bbox{\kappa}}\cdot{\hat  
        {\sf p}}\over m} \tau}  
        e^{-{i\over\hbar}{\bbox{\kappa}}\cdot{\hat  
        {\sf x}}}  
        \langle  
        \lambda  
        |  
        e^{-{i\over\hbar} H_{\text{m}}\tau}  
        b^{\scriptscriptstyle\dagger}_{P+{\kappa\over 2}}  
        b_{P-{\kappa\over 2}}   
        e^{{i\over\hbar} H_{\text{m}}\tau}  
        |  
        \xi  
        \rangle  
        ,  
        \end{eqnarray*}  
to    be inserted in (\ref{aa1}). Before doing this let us  
introduce the useful notation  
        \[  
        e^{-{i\over\hbar} H_{\text{m}}\tau}  
        A  
        e^{{i\over\hbar} H_{\text{m}}\tau}  
        =  
        \sum_\Delta e^{-{i\over\hbar}\Delta t} (A)_\Delta,  
        \quad  
        (A)_\Delta= \sum_E \vert E+\Delta \rangle \langle  
        E+\Delta \vert A\vert E \rangle \langle E\vert,  
        \quad  
        (A)^{\scriptscriptstyle\dagger}_\Delta  
        =  
        (A^{\scriptscriptstyle\dagger})_{- \Delta}  
        .  
        \]  
We have  
        \begin{eqnarray*}  
        {  
        1  
        \over  
         \hbar  
        }  
        \sum^{}_{\xi,\lambda  }  
        {\hat {{\sf L}}}_{\lambda\xi}  {\hat \varrho}  
        {\hat {{\sf L}}}{}_{\lambda\xi}^{\scriptscriptstyle \dagger}  
        &=&  
        {  
        2\varepsilon  
        \over  
                    \hbar^3 {{\sf V}}^2  
        }  
        \sum_{\kappa, P}  
        \sum_{\kappa', P'}  
        \int_0^\infty d\tau \, e^{-{\varepsilon\over\hbar}\tau}  
        e^{{i\over\hbar} {{\bbox{\kappa}}^2\over 2m} \tau}  
        e^{{i\over\hbar} {{\bbox{\kappa}}\cdot{\hat  
        {\sf p}}\over m} \tau}  
        \,  
        {\tilde t} ({\hat  
        {\sf H}}_0+i\varepsilon,{\bbox{\kappa}})  
        \\  
        &&  
        \times  
        \,  
        e^{-{i\over\hbar}{\bbox{\kappa}}\cdot{\hat  
        {\sf x}}}  
        \,  
        {\hat \varrho}  
        \,  
        e^{{i\over\hbar}{\bbox{\kappa}'}\cdot{\hat  
        {\sf x}}}  
        \,  
        {\tilde t}^\dagger ({\hat  
        {\sf H}}_0+i\varepsilon,{\bbox{\kappa}'})  
        \int_0^\infty d\tau' \,  
        e^{-{\varepsilon\over\hbar}\tau'}  
        e^{-{i\over\hbar} {{\bbox{\kappa}'}^2\over 2m} \tau'}  
        e^{-{i\over\hbar} {{\bbox{\kappa}'}\cdot{\hat  
        {\sf p}}\over m} \tau'}  
        \\  
        &&  
        \times  
        \sum_{  
        \Delta ,  
        {\Delta'}  }  
        e^{-{i\over\hbar}\Delta \tau}  
        {\text{Tr}}_{{{{\cal H}_{\scriptscriptstyle F}}}}  
        \left[  
        {  
        {  
        \biggl(  
        b^{\scriptscriptstyle\dagger}_{P+{\kappa\over 2}}  
        b_{P-{\kappa\over 2}}  
        \biggl)_\Delta  
        }        \varrho^{\text{m}}  
        {\biggl(  
        {  
        b^{\scriptscriptstyle\dagger}_{P'-{\kappa'\over 2}}  
        b_{P'+{\kappa'\over 2}}  
        }  
        \biggl)}_{- \Delta'}  
        }  
        \right]  
        e^{{i\over\hbar}{{\Delta'} \tau'}},  
        \end{eqnarray*}  
and similarly for the anticommutator part.  
An important simplification takes place if one can use  
symmetry under time and space translations. Time translation  
invariance occurs if, at least with reference to the  
interaction with the  microsystem, matter can be considered  
at equilibrium, then:  
        \[  
        {\mbox{\rm Tr}}_{{{{\cal H}_{\scriptscriptstyle F}}}}  
        \left(  
        A_\Delta \varrho^{\text{m}}  
        B_{- \Delta'}  
        \right)  
        =  
        \delta_{\Delta,\Delta'}  
        {\mbox{\rm Tr}}_{{{{\cal H}_{\scriptscriptstyle F}}}}  
        \left(  
        A_\Delta \varrho^{\text{m}}  
        B_{-\Delta}  
        \right)  
                  .  
        \]  
Similarly space translation invariance implies  
        \begin{eqnarray*}  
        &&  
        {\text{Tr}}_{{{{\cal H}_{\scriptscriptstyle F}}}}  
        \left[  
        {  
        {\Bigl(  
        b^{\scriptscriptstyle\dagger}_{P+{\kappa\over 2}}  
        b_{P-{\kappa\over 2}}  
        \Bigl)}_\Delta  
        \varrho^{\text{m}}  
        \Bigl(  
        b^{\scriptscriptstyle\dagger}_{P'-{\kappa'\over 2}}  
        b_{P'+{\kappa'\over 2}}  
        \Bigl)_{{- \Delta'}}  
        }  
        \right]=  
        \\  
        &&  
        \hphantom{  
        {\text{Tr}}_{{{{\cal H}_{\scriptscriptstyle F}}}}  
        {  
        {\left(  
        b^{\scriptscriptstyle\dagger}_{P+{\kappa\over 2}}  
        b_{P-{\kappa\over 2}}  
        \right)}_\Delta  
        \varrho^{\text{m}}  
        }  
        }  
        =  
        \delta_{\Delta,\Delta'}  
        \delta_{\kappa,\kappa'}  
        {\text{Tr}}_{{{{\cal H}_{\scriptscriptstyle F}}}}  
        \left[  
        {  
        \Bigl(  
        b^{\scriptscriptstyle\dagger}_{P+{\kappa\over 2}}  
        b_{P-{\kappa\over 2}}   
        \Bigl)_\Delta  
        \varrho^{\text{m}}  
        \Bigl(  
        b^{\scriptscriptstyle\dagger}_{P'-{\kappa\over 2}}  
        b_{P'+{\kappa\over 2}}  
        \Bigl)_{{-\Delta}}  
        }  
        \right]  
        ,  
        \end{eqnarray*}  
such a symmetry can be implemented at equilibrium in the  
thermodynamic limit and can be practically assumed for a  
microsystem interacting with a homogeneous portion of a  
macrosystem. Then one has, performing also the $\tau$,  
$\tau'$ integrals:  
        \begin{eqnarray*}  
        &&  
        \!\!\!\!\!\!  
        \!\!\!\!\!\!  
        \!\!\!\!\!\!  
        \!\!\!\!\!\!  
        {}-{  
        1  
        \over  
          \hbar  
        }  
        \left \{  
        \frac 12  
        \sum^{}_{{\xi,\lambda  }}  
        {\hat {{\sf L}}{}_{\lambda\xi}^{\scriptscriptstyle \dagger}}  
        {\hat {{\sf L}}}_{\lambda\xi} , {\hat \varrho}  
        \right \}  
        +  
        {  
        1  
        \over  
         \hbar  
        }  
        \sum^{}_{{\xi,\lambda  }}  
        {\hat {{\sf L}}}_{\lambda\xi}  {\hat \varrho}  
        {\hat {{\sf L}}{}_{\lambda\xi}^{\scriptscriptstyle \dagger}}  
        =  
        \nonumber   
        \\  
        &&  
        =-{\varepsilon \over \hbar {{\sf V}}^2 }  
        \sum_{\kappa,\Delta}  
        \Biggl(  
        \biggl \{  
        {\hat \varrho},  
        e^{{i\over\hbar}{\bbox{\kappa}}\cdot{{\hat {\sf x}}}}  
        {  
        1  
        \over  
        {{\bbox{\kappa}}\cdot{\hat  
        {\sf p}}\over m}  
        +                 {{\bbox{\kappa}}^2\over 2m}  
        - \Delta -i\varepsilon  
        }  
        {\tilde t}^\dagger ({\hat  
        {\sf H}}_0+i\varepsilon,{\bbox{\kappa}})  
         \nonumber  
        \\  
        &&  
        \hphantom{-{\varepsilon\over \hbar{{\sf V}}^2 }  
        \sum_{\kappa,\Delta}  
        \Biggl(  
        \biggl \{   =  
        }  
        {}  
        \times  
        {\tilde t} ({\hat  
        {\sf H}}_0+i\varepsilon,{\bbox{\kappa}})  
        {  
        1  
        \over  
        {{\bbox{\kappa}}\cdot{\hat  
        {\sf p}}\over m}  
        +                 {{\bbox{\kappa}}^2\over 2m}  
        - \Delta +i\varepsilon  
        }  
        e^{-{i\over\hbar}{\bbox{\kappa}}\cdot{\hat  
        {\sf x}}}  
        \biggl \}   
        \nonumber  
        \\  
        &&  
        \hphantom{-{\varepsilon\over \hbar{{\sf V}}^2}  
        \sum_{\kappa,\Delta}  
        \Biggl(   =  
        }  
        {}  
        -  
        2  
        {  
        1  
        \over  
        {{\bbox{\kappa}}\cdot{\hat  
        {\sf p}}\over m}  
        +                 {{\bbox{\kappa}}^2\over 2m}  
        - \Delta +i\varepsilon  
        }  
        {\tilde t} ({\hat  
        {\sf H}}_0+i\varepsilon,{\bbox{\kappa}})  
        e^{-{i\over\hbar}{\bbox{\kappa}}\cdot{\hat  
        {\sf x}}}  
        \,  
        {\hat \varrho}  
        \,  
        e^{{i\over\hbar}{\bbox{\kappa}}\cdot{{\hat {\sf x}}}}  
        \,  
        {\tilde t}^\dagger ({\hat  
        {\sf H}}_0+i\varepsilon,{\bbox{\kappa}})  
        \nonumber  
        \\  
        &&  
        \hphantom{-{\varepsilon\over \hbar{{\sf V}}^2}  
        \sum_{\kappa,\Delta}  
        \Biggl(      =+  
        }  
        {}        
        \times  
        {  
        1  
        \over  
        {{\bbox{\kappa}}\cdot{\hat  
        {\sf p}}\over m}  
        +                 {{\bbox{\kappa}}^2\over 2m}  
        - \Delta -i\varepsilon  
        }  
        \Biggl)  
        \varrho^{M} (\kappa,\Delta)  
          ,  
        \nonumber  
        \end{eqnarray*}  
where  
        \begin{equation}  
        \label{rodelta}  
        \varrho^{M} (\kappa,\Delta)  
        \equiv  
        {\text{Tr}}_{{{{\cal H}_{\scriptscriptstyle F}}}}  
        \!  
        \left[  
        {\left(            \sum_P  
        b^{\scriptscriptstyle\dagger}_{P+{\kappa\over 2}}  
        b_{P-{\kappa\over 2}}   
        \right)}_{\! \Delta}  
        \varrho^{\text{m}}  
        \left(       \sum_P  
        b^{\scriptscriptstyle\dagger}_{P-{\kappa\over 2}}  
        b_{P+{\kappa\over 2}}  
        \right)_{{\! -\Delta}}  
        \right]  
        ,  
        \end{equation}  
or equivalently, introducing the  
$  
{\hat {\sf x}},  
{\hat {\sf p}}  
$              dependent amplitude  
        \[  
        {\check{t}}({\hat  
        {\sf H}}_0+i\varepsilon,{\bbox{\kappa}},  
        {\hat {\sf x}})  
        =  
        e^{{i\over\hbar}{\bbox{\kappa}}\cdot{\hat  
        {\sf x}}}  
        \,  
        {{t}}({\hat  
        {\sf H}}_0+i\varepsilon,{\bbox{\kappa}}  
        )  
        \,  
        e^{-{i\over\hbar}{\bbox{\kappa}}\cdot{{\hat {\sf x}}}}  
        ,  
        \]  
in the form:  
        \begin{eqnarray}  
        \label{43bis}  
        &&  
        -  
        {  
        1  
        \over  
           \hbar         {{\sf V}}^2  
        }  
        \sum_{\kappa,\Delta}  
        \Biggl(              
        \bigg \{  
        {\hat \varrho},  
        {\check{t}}^\dagger({\hat  
        {\sf H}}_0+i\varepsilon,{\bbox{\kappa}},  
        {\hat {\sf x}})  
        {  
        \varepsilon  
        \over  
        {\left( 
        {{\bbox{\kappa}}\cdot{\hat  
        {\sf p}}\over m}  
        -                 {{\bbox{\kappa}}^2\over 2m}  
        - \Delta  
        \right)}^2 
        +\varepsilon^2  
        }  
        {\check{t}}({\hat  
        {\sf H}}_0+i\varepsilon,{\bbox{\kappa}},  
        {\hat {\sf x}})  
        \bigg \}  
        \nonumber  
        \\  
        &&  
        \hphantom{-{1\over \hbar}  
        {  
        1  
        \over  
                    {{\sf V}}^2  
        }  
        \sum_{\kappa,\Delta}  
        \Biggl(              
        }  
        {}          
        -  
        2\varepsilon  
        \,  
        e^{-{i\over\hbar}{\bbox{\kappa}}\cdot{\hat  
        {\sf x}}}  
        \,  
        {\check{t}}({\hat  
        {\sf H}}_0+i\varepsilon,{\bbox{\kappa}},  
        {\hat {\sf x}})    
        {  
        1  
        \over  
        {{\bbox{\kappa}}\cdot{\hat  
        {\sf p}}\over m}  
        -                 {{\bbox{\kappa}}^2\over 2m}  
        - \Delta +i\varepsilon  
        }  
        {\hat \varrho}  
        \nonumber  
        \\  
        &&  
        \hphantom{-{1\over \hbar}  
        {  
        1  
        \over  
                    {{\sf V}}^2  
        }  
        \sum_{\kappa,\Delta}  
        \Biggl(              
        =}  
        {}          
        \times  
        {  
        1  
        \over  
        {{\bbox{\kappa}}\cdot{\hat  
        {\sf p}}\over m}  
        -                 {{\bbox{\kappa}}^2\over 2m}  
        - \Delta -i\varepsilon  
        }  
        {\check{t}}^\dagger({\hat  
        {\sf H}}_0+i\varepsilon,{\bbox{\kappa}},  
        {\hat {\sf x}})          
        e^{{i\over\hbar}{\bbox{\kappa}}\cdot{{\hat {\sf x}}}}  
        \Biggr)  
        \varrho^{M} (\kappa,\Delta)  
          .  
        \end{eqnarray}  
Introducing this explicit representation in (\ref{n4}) one  
gets the typical master equation of Brownian motion, that  
can be further simplified in the assumption of small  
momentum  transfer, i.e., expanding  the expression with  
respect to  
$  
{{\bbox{\kappa}}\cdot{{\hat {\sf x}}}}  
$  
and  
$  
{{\bbox{\kappa}}\cdot{{\hat {\sf p}}}}  
$.  
Exploiting the fact that $\varrho^{M} (0,\Delta)$ contains a  
$\delta_{\Delta,0}$ factor, one  
can immediately  see by inspection that the $\kappa=0$  
contributions cancel each other provided the effective  
T-matrix is a slow function of energy,  
        \[  
        \left \langle  
        {\bbox{k}}  
        \left |  
        {\hat \varrho}  
        \right |  
                     {\bbox{f}}  
        \right \rangle  
        {\tilde t}(E_k,0)  
        {\tilde t}^\dagger (E_f,0)  
        \approx  
        \left \langle  
        {\bbox{k}}  
        \left |  
        {\hat \varrho}  
        \right |  
                     {\bbox{f}}  
        \right \rangle  \frac 12  
        \left[  
        {\tilde t}^\dagger (E_k,0)  
        {\tilde t}(E_k,0)   +  
        {\tilde t}^\dagger (E_f,0)  
        {\tilde t}         (E_f,0)  
        \right]  
        ;  
        \]  
on the other hand for a homogeneous medium the  
$\kappa=0$  
contributions   
are equal to those obtained writing the correlation function  
as a factorized product  
        \begin{eqnarray*}  
        &&  
        {\text{Tr}}_{{{{\cal H}_{\scriptscriptstyle F}}}}  
        \left[  
        {\left(              \sum_P 
        b^{\scriptscriptstyle\dagger}_{P+{\kappa\over 2}}  
        b_{P-{\kappa\over 2}}   
        \right)}_\Delta 
        \varrho^{\text{m}}  
        {\left(            \sum_P 
        b^{\scriptscriptstyle\dagger}_{P-{\kappa\over 2}}  
        b_{P+{\kappa\over 2}}  
        \right)}_{{-\Delta}} 
        \right]                 \longrightarrow  
        \\  
        &&  
        \hphantom{{\text{Tr}}_{{{{\cal H}_{\scriptscriptstyle F}}}}  
        {\left(              \sum_P 
        b^{\scriptscriptstyle\dagger}_{P+{\kappa\over 2}}  
        b_{P-{\kappa\over 2}}   
        \right)}_\Delta 
        \varrho^{\text{m}}  
        }  
        {\text{Tr}}_{{{{\cal H}_{\scriptscriptstyle F}}}}  
        \Biggl[  
        {\left(       \sum_P 
        b^{\scriptscriptstyle\dagger}_{P+{\kappa\over 2}}  
        b_{P-{\kappa\over 2}}   
        \right)}_\Delta 
        \varrho^{\text{m}}  
        \Biggr]  
        {\text{Tr}}_{{{{\cal H}_{\scriptscriptstyle F}}}}  
        \Biggl[  
        {\left(        \sum_P 
        b^{\scriptscriptstyle\dagger}_{P-{\kappa\over 2}}  
        b_{P+{\kappa\over 2}}  
        \right)}_{{-\Delta}} 
        \varrho^{\text{m}}  
        \Biggr]  
            ,  
        \end{eqnarray*}  
provided we assume the condition of  ``normal density fluctuations'',  
$  
(  
\left \langle  
N^2  
\right \rangle  -  
{\left \langle 
N  
\right \rangle}^2 
)  /   
V^2  \ll n_{\text{o}}^2 
$.  
Instead of restricting the sum to the $\kappa \not= 0$  
contributions  we can therefore subtract from the  
correlation function its factorized part. After  
straightforward manipulations, using  
        \[  
        \sum_P  
        b^{\scriptscriptstyle\dagger}_{P+{\kappa\over 2}}  
        b_{P-{\kappa\over 2}}  
        =  
        {\int d^3 \! {\bbox{x}} \,}  
        \psi^{\scriptscriptstyle\dagger}({\bbox{x}})  
        \psi({\bbox{x}})  
        e^{{i\over\hbar}{\bbox{\kappa}}\cdot{\bbox{x}}},  
        \qquad  
        {\text{Tr}}_{{{{\cal H}_{\scriptscriptstyle F}}}}  
        \left[  
        {\left(       A               \right)}_\Delta 
        \varrho^{\text{m}}  
        {\left( 
        B  
        \right)}_{{-\Delta}} 
        \right]  
        =  
        \int  
        {  
        dt  
        \over  
          2\pi\hbar  
        }  
        e^{  
        -{i\over\hbar} \Delta t  
        }  
        \,  
        \left \langle  
        B A (t)  
        \right \rangle  
        \]  
we come to  
        \begin{eqnarray*}  
        &&  
        {\text{Tr}}_{{{{\cal H}_{\scriptscriptstyle F}}}}  
        \left[  
        {\left(              \sum_P 
        b^{\scriptscriptstyle\dagger}_{P+{\kappa\over 2}}  
        b_{P-{\kappa\over 2}}   
        \right)}_\Delta 
        \varrho^{\text{m}}  
        {\left(            \sum_P 
        b^{\scriptscriptstyle\dagger}_{P-{\kappa\over 2}}  
        b_{P+{\kappa\over 2}}  
        \right)}_{{-\Delta}} 
        \right]  
        \\  
        &&  
        {}  
        -  
        {\text{Tr}}_{{{{\cal H}_{\scriptscriptstyle F}}}}  
        \Biggl[  
        {\left(       \sum_P 
        b^{\scriptscriptstyle\dagger}_{P+{\kappa\over 2}}  
        b_{P-{\kappa\over 2}}   
        \right)}_\Delta 
        \varrho^{\text{m}}  
        \Biggr]  
        {\text{Tr}}_{{{{\cal H}_{\scriptscriptstyle F}}}}  
        \Biggl[  
        {\left(        \sum_P 
        b^{\scriptscriptstyle\dagger}_{P-{\kappa\over 2}}  
        b_{P+{\kappa\over 2}}  
        \right)}_{{-\Delta}} 
        \varrho^{\text{m}}  
        \Biggr]        =  
        \\  
        &&  
        \hphantom{  
        {}  
        -  
        {\text{Tr}}_{{{{\cal H}_{\scriptscriptstyle F}}}}  
        \left[  
        {\left(       \sum_P 
        b^{\scriptscriptstyle\dagger}_{P+{\kappa\over 2}}  
        b_{P-{\kappa\over 2}}   
        \right)}_\Delta 
        \varrho^{\text{m}}  
        \right]  
        }  
        =  
        \int  
        {  
        dt  
        \over  
          2\pi\hbar  
        }  
        e^{  
        -{i\over\hbar} \Delta t  
        }  
        {\int d^3 \! {\bbox{x}} \,}  
        {\int d^3 \! {\bbox{y}} \,}  
        e^{{i\over\hbar}{\bbox{\kappa}}\cdot({\bbox{x}}-{\bbox{y}})}  
        \left \langle  
        \delta N({\bbox{y}})  
        \delta N({\bbox{x}},t)  
        \right \rangle  
        \end{eqnarray*}  
where  
        \[  
        \left \langle  
        N({\bbox{x}})  
        \right \rangle  
        =  
        {\text{Tr}}_{{{{\cal H}_{\scriptscriptstyle F}}}}  
        \left[  
        N({\bbox{x}})  
        \varrho^{\text{m}}  
        \right]   ,  
        \qquad  
        \delta N({\bbox{x}}) =N({\bbox{x}})  -  
        \left \langle  
        N({\bbox{x}})  
        \right \rangle    ,  
        \]  
and finally:  
        \begin{eqnarray}  
        \label{aa3}  
        &&  
        \!\!\!\!\!\!  
        \!\!\!\!\!\!  
        \!\!\!\!\!\!  
        {}-{  
        1  
        \over  
          \hbar  
        }  
        \left \{  
        \frac 12  
        \sum^{}_{{\xi,\lambda  }}  
        {\hat {{\sf L}}{}_{\lambda\xi}^{\scriptscriptstyle \dagger}}  
        {\hat {{\sf L}}}_{\lambda\xi} , {\hat \varrho}  
        \right \}  
        +  
        {  
        1  
        \over  
         \hbar  
        }  
        \sum^{}_{{\xi,\lambda  }}  
        {\hat {{\sf L}}}_{\lambda\xi}  {\hat \varrho}  
        {\hat {{\sf L}}{}_{\lambda\xi}^{\scriptscriptstyle \dagger}}  
        =  
        \nonumber \\  
        &&  
        =  
        {}-{\varepsilon \over \hbar {{\sf V}}^2}  
        \sum_{\kappa,\Delta}  
        \Biggl(  
        \biggl \{  
        {\hat \varrho},  
        e^{{i\over\hbar}{\bbox{\kappa}}\cdot{{\hat {\sf x}}}}  
        {  
        1  
        \over  
        {{\bbox{\kappa}}\cdot{\hat  
        {\sf p}}\over m}  
        +                 {{\bbox{\kappa}}^2\over 2m}  
        - \Delta -i\varepsilon  
        }  
        {\tilde t}^\dagger ({\hat  
        {\sf H}}_0+i\varepsilon,{\bbox{\kappa}})  
        \nonumber  
        \\  
        &&  
        \hphantom{=  
        {}-{\varepsilon \over \hbar {{\sf V}}^2}  
        \sum_{\kappa,\Delta}  
        \Biggl(   =  
        }  
        {}  
        \times  
        {\tilde t} ({\hat  
        {\sf H}}_0+i\varepsilon,{\bbox{\kappa}})  
        {  
        1  
        \over  
        {{\bbox{\kappa}}\cdot{\hat  
        {\sf p}}\over m}  
        +                 {{\bbox{\kappa}}^2\over 2m}  
        - \Delta +i\varepsilon  
        }  
        e^{-{i\over\hbar}{\bbox{\kappa}}\cdot{\hat  
        {\sf x}}}  
        \biggr \}  
        \nonumber  
        \\  
        &&  
        \hphantom{=  
        {}-{\varepsilon \over \hbar {{\sf V}}^2}  
        \sum_{\kappa,\Delta}  
        \Biggl(  
        }   
        -  
        2  
        {  
        1  
        \over  
        {{\bbox{\kappa}}\cdot{\hat  
        {\sf p}}\over m}  
        +                 {{\bbox{\kappa}}^2\over 2m}  
        - \Delta +i\varepsilon  
        }  
        {\tilde t} ({\hat  
        {\sf H}}_0+i\varepsilon,{\bbox{\kappa}})  
        e^{-{i\over\hbar}{\bbox{\kappa}}\cdot{\hat  
        {\sf x}}}  
        \,  
        {\hat \varrho}  
        \nonumber  
        \\  
        &&  
        \hphantom{=  
        {}-{\varepsilon \over \hbar {{\sf V}}^2}  
        \sum_{\kappa,\Delta}  
        \Biggl(  
        \biggl \{  
        =     
        }   
        \times     
        e^{{i\over\hbar}{\bbox{\kappa}}\cdot{{\hat {\sf x}}}}  
        \,  
        {\tilde t}^\dagger ({\hat  
        {\sf H}}_0+i\varepsilon,{\bbox{\kappa}})  
         {  
        1  
        \over  
        {{\bbox{\kappa}}\cdot{\hat  
        {\sf p}}\over m}  
        +                 {{\bbox{\kappa}}^2\over 2m}  
        - \Delta -i\varepsilon  
        }  
        \Biggr)                          
        \nonumber  
        \\  
        &&  
        \hphantom{=  
        {}-{\varepsilon \over \hbar {{\sf V}}^2}  
        \sum_{\kappa,\Delta}  
        }   
        \times     
        \int  
        {  
        dt  
        \over  
          2\pi\hbar  
        }  
        e^{  
        -{i\over\hbar} \Delta t  
        }  
        {\int d^3 \! {\bbox{x}} \,}  
        {\int d^3 \! {\bbox{y}} \,}  
        e^{{i\over\hbar}{\bbox{\kappa}}\cdot{\bbox{x}}}  
        \left \langle  
        \delta N({\bbox{y}})  
        \delta N({\bbox{x}}+{\bbox{y}},t)  
        \right \rangle  
           .  
        \end{eqnarray}  
Thanks to the last term of (\ref{n4}) it is possible to take  
into account collisions that modify  
the state of the  macroscopic system  
(see~\cite{art1}).  
The probability per unit time of such collisions is given by  
the trace of  
$  
        {  
        1  
        \over  
         \hbar  
        }  
        \sum^{}_{{\xi,\lambda  }}  
        {\hat {{\sf L}}}_{\lambda\xi}  {\hat \varrho}  
        {\hat {{\sf L}}{}_{\lambda\xi}^{\scriptscriptstyle \dagger}}  
$  
as seen in Sec.~II.
In the case considered  
this trace may be written as  
        \begin{eqnarray}  
        \label{copia}  
        &&  
        {  
        2\pi  
        \over  
            \hbar  
        }  
        {  
        n_o  
        \over  
           (2\pi\hbar)^4  
        }  
        {\int d^3 \! {\bbox{k}} \,}  
        {\int d^3 \! {\bbox{\kappa}} \,}  
        \left \langle  
        {\bbox{\kappa}}  
        \left |  
              {\hat \varrho}  
        \right |  
        {\bbox{\kappa}}  
        \right \rangle  
        {\left | 
        {\tilde t}(E_k, {\bbox{\kappa}}-{\bbox{k}})  
        \right |}^2 
        \\  
        &&  
        \hphantom{        -  
        {  
        2\pi  
        \over  
            \hbar  
        }  
        {  
        n_o  
        \over  
           (2\pi\hbar)^4  
        }  
        {\int d^3 \! {\bbox{k}} \,}  
        {\int d^3 \! {\bbox{\kappa}} \,}  
        }  
        \times  
        \int dt \,  
        {\int d^3 \! {\bbox{x}} \,}  
        e^{  
        -{i\over\hbar}  
        \left(  
        {  
        {\bbox{\kappa}}^2  
        \over  
                         2m  
        }                  -  
        {  
        {\bbox{k}}^2  
        \over  
                    2m  
        }  
        \right)       t  
        +{i\over\hbar}  
        ({\bbox{\kappa}}-{\bbox{k}})\cdot{\bbox{x}}  
        }  
        {\int d^3 \! {\bbox{y}} \,}  
        {  
        1  
        \over  
         N  
        }  
        \left \langle  
        \delta N({\bbox{y}})  
        \delta N({\bbox{x}}+{\bbox{y}},t)  
        \right \rangle  
           ,  
        \nonumber  
        \end{eqnarray}  
thus recovering again the van Hove structure for the  
scattering cross section [compare (\ref{wigner})], with the  
difference that now the system is considered to be  
homogeneous, so that only the momentum distribution of the  
incoming  microsystem is of relevance. Let us observe that  
subtraction of the uncorrelated part of the response  
function accounts for the fact that only diffuse scattering,  
that is scattering that does not leave the macroscopic  
system unchanged~\cite{Sears}, contributes to this term. We  
now specialize to the case of neutrons, adopting the Fermi  
pseudopotential given by (\ref{fermi}), so that  
(\ref{copia}) becomes  
        \begin{equation}  
        \label{aa2}  
        {  
        1  
        \over  
            \hbar  
        }  
        n_o  
        {  
        b^2  
        \over  
           m^2  
        }  
        {\int d^3 \! {\bbox{k}} \,}  
        {\int d^3 \! {\bbox{\kappa}} \,}  
        \left \langle  
        {\bbox{\kappa}}  
        \left |  
              {\hat \varrho}  
        \right |  
        {\bbox{\kappa}}  
        \right \rangle  
        S_{\text{c}}  
        \left(  
        {  
        {1\over\hbar}  
        \left[  
        {\bbox{\kappa}}-{\bbox{k}}  
        \right]               ,  
        {1\over\hbar}  
        \left[  
        {  
        {\bbox{\kappa}}^2  
        \over  
                         2m  
        }                  -  
        {  
        {\bbox{k}}^2  
        \over  
                    2m  
        }  
        \right]  
        }  
        \right)  
        ,  
        \end{equation}  
where, denoting by $\omega$ and ${\bbox{q}}$ energy and  
momentum transfer respectively,  
        \begin{equation}  
        \label{respf}  
        S_{\text{c}} ({\bbox{q}},\omega)=  
        {  
        1  
        \over  
         2\pi N  
        }  
        \int dt \,  
        {\int d^3 \! {\bbox{x}} \,}  
        e^{  
        -i(\omega t - {\bbox{q}}\cdot{\bbox{x}})  
        }  
        {\int d^3 \! {\bbox{y}} \,}  
        \left \langle  
        \delta N({\bbox{y}})  
        \delta N({\bbox{x}}+{\bbox{y}},t)  
        \right \rangle  
        .  
        \end{equation}  
If the momentum distribution of the incoming particle is  
suitably peaked around ${\bbox{p}_0}$ with respect to the  
momentum dependence of $S_{\text{c}}$ we have from (\ref{aa2})  
        \[  
        { n_o  
        b^2  
        \over  
        \hbar   m^2  
        }  
        {\int d^3 \! {\bbox{k}} \,}  
        {\int d^3 \! {\bbox{\kappa}} \,}  
        \left \langle  
        {\bbox{\kappa}}  
        \left |  
              {\hat \varrho}  
        \right |  
        {\bbox{\kappa}}  
        \right \rangle  
        S_{\text{c}}  
        \left(  
        {1\over\hbar}  
        \left[  
        {\bbox{p}_0}-{\bbox{k}}  
        \right]               ,  
        \omega_{p_0}-\omega_k  
        \right)  
        =  
        {  n_o        b^2  
        \over  
        \hbar   m^2  
        }  
        {\int d^3 \! {\bbox{k}} \,}  
        S_{\text{c}}  
        \left(  
        {1\over\hbar}  
        \left[  
        {\bbox{p}_0}-{\bbox{k}}  
        \right]               ,  
        \omega_{p_0}-\omega_k  
        \right)  
        \]  
in particular, in the static limit expression (\ref{aa2})  
becomes:  
        \[  
         n_o b^2 {p_0 \over m}  
        \int d\Omega_q  \, S_{\text{c}} ({\bbox{q}})=  
        n_o {p_0 \over m} \sigma_{\text{d}}  
        \]  
where  
        \[  
        S_{\text{c}} ({\bbox{q}})  
        =  
        {  
        1  
        \over  
          N  
        }  
        {\int d^3 \! {\bbox{x}} \,}  
        e^{  
        i{\bbox{q}}\cdot{\bbox{x}}  
        }  
        {\int d^3 \! {\bbox{y}} \,}  
        \left \langle  
        \delta N({\bbox{y}})  
        \delta N({\bbox{x}}+{\bbox{y}})  
        \right \rangle         ,  
        \]  
and we have denoted by ${\bbox{q}}$ the momentum  
transfer and by $\sigma_{\text{d}}$ the total diffusion cross  
section per particle. This is the result derived by Sears for the  
attenuation of the  coherent beam due to incoherent  
scattering, which he obtains by an evaluation of the local  
field effects, neglected in the equation giving the optical neutron  
dynamics  (\ref{Sears}), see~\cite{Sears,21,Searsphysica}. In our  
approach, however, the incoherent  
contribution is already present in the equation giving the  
dynamics of the  microsystem,  being connected to the  
thermodynamic properties of the  macrosystem  through the  
response function $S_{\text{c}}({\bbox{q}},\omega)$. This  
new feature is obtained by means of the more general  
formalism adopted, leading to a master-equation of the  
Lindblad type for the  statistical operator, in which due to  
the optical theorem a close correlation exists between the  
incoherent  contribution and the imaginary part of the  
optical potential which is not connected to absorption. To  
see this correction to the optical potential let us exploit  
the simple relation  
        \[  
        {\hat {\sf A}}  
        = {\hat {\sf A}}^\dagger,  
        \quad  
        {\hat {{\sf B}}}  
        = {\hat {{\sf B}}}^\dagger,  
        \quad  
        {\hat {{\sf U}}}=  
        {\hat {\sf A}} +i  
        {\hat {{\sf B}}}  
        \qquad  
        \Rightarrow  
        \qquad  
        {\hat {{\sf U}}}  
        {\hat \varrho}-  
        {\hat \varrho}  
        {\hat {{\sf U}}}^\dagger   =  
        \left[  
        {\hat {\sf A}}   ,  
        {\hat \varrho}  
        \right]              +i  
        \left \{  
        {\hat {{\sf B}}}   ,  
        {\hat \varrho}  
        \right \}  
        \]  
and write the commutator and anticommutator term of  
(\ref{n4}) in the form  
$-{i\over\hbar}  
\left(  
        {\hat {{\sf U}}}  
        {\hat \varrho}-  
        {\hat \varrho}  
        {\hat {{\sf U}}}^\dagger  
\right)  
$.  
The calculation of ${\hat {{\sf U}}}$ is essentially given  
by the anticommutator at the r.h.s. of (\ref{aa3}) and the  
commutator in  (\ref{b8}). In the case of the Fermi  
pseudopotential, using   (\ref{respf}) one has:  
        \begin{equation}  
        \label{aa5}  
        {\hat {{\sf U}}}  =  
        {  
        2\pi \hbar^2  
        \over  
                    m  
        }  
        n_o  
        \left[  
        b -i  
        {  
        b^2  
        \over  
           4\pi  
        }  
        {\int d^3 \! {\bbox{k}} \,}  
        \left |  
        {\bbox{k}}  
        \rangle\langle  
        {\bbox{k}}  
        \right |  
        \int d\omega_\kappa \,  
        \int d\Omega_\kappa \,  
                {  
        \kappa  
        \over  
         \hbar  
                }  
        \,  
        S_{\text{c}}  
        \left(  
        {1\over\hbar}  
        \left[  
        {\bbox{\kappa}}-{\bbox{k}}  
        \right]               ,  
        {1\over\hbar}  
        \left[  
        {  
        {\bbox{\kappa}}^2  
        \over  
                         2m  
        }                  -  
        {  
        {\bbox{k}}^2  
        \over  
                    2m  
        }  
        \right]  
        \right)  
        \right]  
        ,  
        \end{equation}  
or in the static limit  
        \begin{equation}  
        \label{aa4}  
        {\hat {{\sf U}}}  =  
        {  
        2\pi \hbar^2  
        \over  
                    m  
        }  
        n_o  
        \left[  
        b -i  
        {  
        b^2  
        \over  
           4\pi  
        }  
        {\int d^3 \! {\bbox{k}} \,}  
        \left |  
        {\bbox{k}}  
        \rangle\langle  
        {\bbox{k}}  
        \right |  
        {k\over \hbar}  
        \int d\Omega_q \,  
        S_{\text{c}}({\bbox{q}})  
        \right]  
        ,  
        \end{equation}  
where ${\bbox{q}}$ denotes as usual the momentum transfer.  
Neglecting diffuse scattering we would have  
$  
{\hat {{\sf U}}}=  
{  
2\pi \hbar^2  
\over  
            m  
}n_o b$, simply a c-number giving the usual refractive  
index; the remaining part is, in a sense, induced by  
the optical theorem.  
To compare with the results derived by Sears we have to  
consider the expression obtained for the static limit  
(\ref{aa4}) applied to a plane wave of momentum  
${\bbox{p}_0}$, which gives an idealized description of the  
preparation of the incoming  microsystem, thus leading to  
        \begin{equation}  
        \label{correz}  
        {\hat {{\sf U}}}  =  
        {  
        2\pi \hbar^2  
        \over  
                    m  
        }  
        n_o  
        \left[  
        b -i  
        {  
        b^2  
        \over  
           4\pi  
        }  
        {p_0\over \hbar}  
        \int d\Omega_q \,  
        S_{\text{c}}({\bbox{q}})  
        \right]  
            ;  
        \end{equation}  
this expression agrees with the results obtained relying on  
the idea of local field corrections (see~\cite{Sears}, Ch.  
4), however here (\ref{aa4}) is a direct consequence of the equation  
driving the dynamics and of the Ansatz  (\ref{tpheno}).  
The analysis that we have put forward relies on the  
assumption that the main contribution to the dynamics is  
given by the commutator term in  (\ref{n4}), while the terms  
in  (\ref{aa1}) may, as a first approximation, be neglected.  
This leads to an optical description, as for the case of  
neutrons, in which, considering the  
dimensionless parameter ${2\pi \hbar^2 \over mE}n_{\text{o}}  
b$, the terms other than the commutator are of second order.  
The opposite situation takes place if the interaction is  
such that the main contribution is given by  (\ref{aa1}),  
while the commutator may be neglected. This happens when  
dissipative effects are predominant, as in the case of  
Brownian motion mentioned below (\ref{43bis}), where  
incoherent interactions through  
collisions involving energy and momentum transfer play the  
main role, a case we intend to deal with in a future paper.  
\section{EXPERIMENTAL IMPLICATIONS}  
We now address our attention to potential experimental  
implications of the above introduced description of  
neutron-matter interaction. Of course possible new features  
in the dynamics are linked to the presence of the last two  
terms in the r.h.s. of (\ref{n4}), as given by (\ref{aa3}),  
and such corrections will  
be generally small, being of second order  
in ${2\pi \hbar^2 \over mE}n_{\text{o}}b$  
(typically ${2\pi \hbar^2 \over mE}n_{\text{o}}b \leq  
10^{-5}$ at thermal neutron energies).  
In this respect interferometric experiments,  
in which the experimental setup is  
conceived in order to enhance the  coherent behavior, should be  
particularly relevant: think for example of the beautiful  
experiments realized  
by the Rauch group in Wien exploiting the perfect crystal  
neutron interferometer~\cite{Erice,6,Rauch}.  
\par  
Consider now  
eq.(\ref{n4}): the map on the r.h.s. is affine and trace  
preserving, and therefore clearly predicts neutron  
conservation. Nevertheless  
the last  contribution which offsets the anticommutator  
term is linked to diffuse scattering: one has  
neutron conservation if also diffuse particles contribute  
to the experimental observation. This is not so for  
interferometric experiments. In such cases only the wavelike  
behavior  affects the observed dynamics, and thus  
only the commutator part of the evolution map is of  
relevance: the net result is an imaginary correction to the  
coherent scattering length as in (\ref{correz}), that is to  
say a reduction of the neutron flux responsible for the  
interference pattern.  
This fact is usually taken into account adding an imaginary  
part proportional to the total scattering cross-section  
$\sigma_{\text{t}}$ to the phase shift calculated as in  
(\ref{phaseshift}), thus including both absorption and  
diffuse scattering (see~\cite{6,7})  
according to the formula:  
        \[  
        \chi  
        =  
        {\chi}^{'}+i  
        {\chi}^{''}  
        =  
        - n_{\text{o}} b \lambda D + i n_{\text{o}} \sigma_{\text{t}}  
        {D \over 2}  
        ,  
        \qquad  
        \exp({i \chi})  
        =  
        \exp  
        \left(  
        {-i n_{\text{o}} b \lambda D -n_{\text{o}} \sigma_{\text{t}}  
        {D \over 2} }  
        \right)  
              .  
        \]  
In the absence of absorption this  
correction is considered negligible and the relevant  
incident flux is often evaluated simply closing one of the two  
beam paths. This attitude is however at least in principle incorrect,
as it appears taking 
the whole dynamics as given by (\ref{n4}) into account. In  
fact when one closes the path without the sample also  
diffuse neutrons, which are lost for the interference  
pattern, having their path ``labeled'' by scattering with the  
sample, may contribute to the transmitted intensity. The  
experimental device is no more acting as an interferometer  
and therefore cannot select only those neutrons that have  
undergone coherent interactions. This additional  
contribution to the transmitted neutron flux is given by the  
trace of the last term of (\ref{n4}), that is to say by  
(\ref{aa2}). In calculating the amplitude of the  
interference pattern one should therefore rely not simply on  
the measured transmitted flux, but on this quantity minus  
the additional incoherent  contribution  given by  
(\ref{aa2}), thus obtaining a reduction of this amplitude:
the purely ``optical'' treatment leads in principle to an overestimate
of the visibility of the interference pattern.
This is normally not the case in real experiments, since the angle of  
acceptance of diffuse neutrons is  very small, as for the perfect crystal  
neutron interferometer. Let us give some  
quantitative estimate of the aforementioned effect.  
\par  
In order to evaluate (\ref{aa2}) we have to make a definite  
choice for the structure function  
$S_{\text{c}} ({\bbox{q}},\omega)$, in fact   
(\ref{aa2}) is given by:  
        \begin{eqnarray*}  
        &&  
        {\cal A} \equiv  
        {  
        1  
        \over  
         \hbar  
        }  
        {\text{Tr}}_{{\cal H}^{(1)}}  
        \sum^{}_{{\xi,\lambda  }}  
        {\hat {{\sf L}}}_{\lambda\xi}  {\hat \varrho}  
        {\hat {{\sf L}}{}_{\lambda\xi}^{\scriptscriptstyle \dagger}}  
        =  
        {  
        n_o  
        b^2  
        \over  
        \hbar           m^2  
        }  
        {\int d^3 \! {\bbox{k}} \,}  
        {\int d^3 \! {\bbox{\kappa}} \,}  
        \left \langle  
        {\bbox{\kappa}}  
        \left |  
              {\hat \varrho}  
        \right |  
        {\bbox{\kappa}}  
        \right \rangle  
        S_{\text{c}}  
        \left(  
        {  
        {1\over\hbar}  
        \left[  
        {\bbox{\kappa}}-{\bbox{k}}  
        \right]               ,  
        {1\over\hbar}  
        \left[  
        {  
        {\bbox{\kappa}}^2  
        \over  
                         2m  
        }                  -  
        {  
        {\bbox{k}}^2  
        \over  
                    2m  
        }  
        \right]  
        }  
        \right)  ,  
        \end{eqnarray*}  
where the quantity ${\cal A}$ takes diffusion at any angle into  
account.  
In the static approximation, for a homogeneous and isotropic  
medium, such as a liquid or a gas, one has~\cite{Sears}:  
        \begin{eqnarray}  
        \label{simple}  
        &&  
        S_{\text{c}} ({\bbox{q}},\omega)  
        =  
        S_{\text{c}} ({\bbox{q}}) \delta(\omega)  
        \qquad \qquad  
        S_{\text{c}} ({\bbox{q}})  
        =  
        1  
        +  
        n_{\text{o}}  
        \int \! d^3 \! {\bbox{r}} \,  
        e^{i {\bbox{q}}\cdot{\bbox{r}}}  
        [g(r)-1]   ,  
        \end{eqnarray}  
where $g(r)$ is the pair correlation function. A possible  
choice for $g(r)$, allowing  $S_{\text{c}} ({\bbox{q}})$ to  
be evaluated analitically, is the following, valid for a  
dilute hard sphere gas with atomic diameter $a$:  
        \[  
        g(r)=  
        \left \{  
        \begin{array}{ll}  
        0 & \quad r<a  
        \\  
        1 & \quad r>a  
        \end{array}  
        \right.  .  
        \]  
The quantity of interest for us is ${\cal A}$ in its  
dependence from the maximal angular acceptance $\varphi$,  
determined by the experimental apparatus, multiplied by the  
time the neutron takes to go through the sample. Supposing  
the momentum distribution of the incoming particle  
sufficiently  
well peaked around ${\bbox{p}_0}$ we rewrite ${\cal A}$ 
introducing the expression given by (\ref{simple}) and  
multiplying by the time interval, thus coming to:  
        \begin{eqnarray*}  
        {\cal A}(\varphi)  
        &=&  
        2 \pi n_{\text{o}} b^2 D  
        \int_0^{\varphi} \! d\theta \sin \theta  
        \\  
        &&  
        {}  
        \times  
        \left \{  
        1  
        -  
        {  
        2 \pi n_{\text{o}} a^3  
        \over  
        (1 - \cos \theta)  
        }  
        {  
        {\left( 
        {  
        \hbar  
        \over  
        a p_0  
        }  
        \right)}^2 
        }  
        \left[  
        {  
        \sin  
        \left(  
        {  
        a p_{\text{o}}  
        \over  
                \hbar  
        }  
        \sqrt{2(1 - \cos \theta) }  
        \right)  
        \over  
        {  
        a p_{\text{o}}  
        \over  
                \hbar  
        }  
        \sqrt{2(1 - \cos \theta) }  
        }  
        -  
        \cos  
        \left(  
        {  
        a p_{\text{o}}  
        \over  
                \hbar  
        }  
        \sqrt{2(1 - \cos \theta) }  
        \right)        \right]  
        \right \}  
        ,  
        \end{eqnarray*}  
where  
$  
\cos \theta  
=  
({\bbox{p}_0}   \cdot  {\bbox{k}})  
/  
p_0^2  
$.  
The primitive of this integral can be straightforwardly evaluated 
by a change of variables, 
and exploiting the fact that in our  
model  
$  
S_{\text{c}}    (0)  
=  
1  
-  
{4\over 3}\pi a^3 n_{\text{o}}  
$, we have an explicit representation of 
diffuse scattering at any angle $\varphi$:  
        \begin{eqnarray*}
        {\cal A}(\varphi)  
        &=&  
        2 \pi n_{\text{o}} b^2 D  
        \left \{  
        (1 - \cos \varphi)  
        +  
        3  
        [1 - S_{\text{c}}    (0) ]  
        {  
        {\left( 
        {  
        \hbar  
        \over  
        a p_0  
        }  
        \right)}^2 
        }  
        \left[  
        {  
        \sin  
        \left(  
        {  
        a p_{\text{o}}  
        \over  
                \hbar  
        }  
        \sqrt{2(1 - \cos \varphi) }  
        \right)  
        \over  
        {  
        a p_{\text{o}}  
        \over  
                \hbar  
        }  
        \sqrt{2(1 - \cos \varphi) }  
        }  
        -  
        1  
        \right]  
        \right \}  
        ;  
        \end{eqnarray*}  
considering in particular small $\varphi$ the expression may be 
approximated  
as:  
        \begin{eqnarray*}  
        {\cal A}(\varphi)  
        &\simeq&  
        \pi n_{\text{o}} b^2 D  
        \left \{  
        {\varphi}^2  
        S_{\text{c}}    (0)  
        +  
        {\varphi}^4  
        {\left(  
        {1\over 20}  
        [1 - S_{\text{c}}    (0) ]  
        {{\left( 
        {  
        a p_0  
        \over  
        \hbar
        }  
        \right)}^2} 
        -  
        {1\over 12}  
        S_{\text{c}}    (0)  
        \right)}  
        +  
        O({\varphi}^6)  
        \right \}  
        .  
        \end{eqnarray*}  
\par  
Let us now consider the experiments performed using the  
perfect crystal interferometer. The angular acceptance is  
very small, only a few microradians for thermal  
neutrons~\cite{5}. Taking for instance a gaseous  
sample, an order  of magnitude estimate gives  
${\cal A}(\varphi)\simeq10^{-14}$, that is to say an extremely small  
quantity, in agreement with the accuracy obtained using this  
interferometer based on Bragg diffraction. An interferometer  
based on a different physical principle could possibly lead  
to a higher angular acceptance, thus enhancing this effect  
connected to diffusion.  
In view of the next equation (\ref{critico}) 
a completely different situation arises if one considers 
systems with abnormally large density fluctuations, 
as would be the case near a first order 
phase-transition.  
\par  
Another point of interest is the  
linear dependence on  
$S_{\text{c}}(0)$  
of the leading term in ${\cal A}(\varphi)$.  
The quantity $S_{\text{c}}(0)$ is particularly relevant from  
the physical point of view, being connected to the  
isothermal compressibility $\chi_{\text{T}}$ and to the  
fluctuations in the number of particles in the  
sample~\cite{Egelstaff}:  
        \begin{equation}  
        \label{critico}  
        S_{\text{c}}(0)  
        =  
        n_{\text{o}}  
        k_{\text{B}}   T  
        \chi_{\text{T}}  
        =  
        \overline{  
        {  
        (\Delta N)^2  
        \over  
        N  
        }  
        }  
        .  
        \end{equation}  
The actual value of $S_{\text{c}}(0)$  cannot be measured  
experimentally from scattering experiments, and has to be  
obtained by an analytical continuation. The analysis we  
propose could provide an independent way to measure  
$S_{\text{c}}$ at ${\bbox{q}}=0$. In fact in the static  
approximation, independently of the particular form  
of $S_{\text{c}}({\bbox{q}})$, for very small  
$| {\bbox{q}} |$, that is to say for very small $\varphi$,  
one has   in good approximation:  
        \[  
        {\cal A}(\varphi)\simeq \pi n_{\text{o}} b^2 D\, S_{\text{c}}(0)  
        {\varphi}^2  
        .  
        \]  
The value of $S_{\text{c}}(0)$ could then  be obtained, at  
least in principle, comparing the amplitude of the  
interference pattern with the measured transmitted  
intensity.  
\section{SUMMARY AND OUTLOOK}  
The example of neutron interaction with matter 
has been discussed inside the approach outlined in~\cite{art1,japan,berlin} 
to describe the  
subdynamics of a   microsystem interacting with a system  
having  
many degrees of freedom. The formal scheme leads to a generator for  
the irreversible time evolution of the Lindblad form, whose  
expression relies on suitable choices for the potential term  
related to the T-matrix and the  statistical operator describing the  
thermodynamic state of the system. In the example considered the  
main ingredient is given by the Fermi pseudopotential adopted to  
describe the neutron-nucleus interaction in impulse  
approximation. Then we obtain from  
(\ref{Lind}), neglecting the  incoherent  contribution, the  
equation used by Sears to describe all neutron optical phenomena,  
as well as known expressions for the index of refraction.  
The  incoherent  contribution is  
necessary to fulfill the optical theorem and take diffuse  
scattering, that attenuates the  coherent beam, into account. We  
have also shown how it may be connected to properties of the  
macrosystem, as expressed by the dynamic structure function.  
Furthermore possible experimental implications have been  
discussed in Sec.~V.  
\par  
Even though it introduces a smaller correction the  incoherent  
contribution is very important from the theoretical point of  
view. We expect that it will help studying the tricky borderline  
between  a pure optical wavelike behavior and the fully  
incoherent particlelike one, based on a diffusion equation: 
in fact~(\ref{n4}) 
leads in a direct way to the theory of Brownian motion , as stressed 
under~(\ref{43bis}); a somewhat similar treatment, in the case 
of an ideal gas, 
has been given in~\cite{Diosi}. 
It is not surprising that
the incoherent contribute to the dynamics  
has grown out of a thoroughly quantum mechanical  
treatment, as shown by the typical quantum structure of the  
Lindblad equation, relying on non-commutating   
operators, in which an essential role is played by the  
statistical operator $\varrho$, rather then by  
the wave function $\psi$. This point is of central relevance,  
since the  terms which describe the  incoherent  
dynamics cannot be introduced in the  formalism of the wave  
function and are therefore unavoidably absent in an optical-like  
treatment, simply reminiscent of classical optical descriptions.  
\par  
We hope that this study of the emergence of incoherence in 
neutron-matter interaction
will lead to a better understanding of the general problem of 
irreversibility 
and of description of non-equilibrium systems. Typically 
coexistence of an 
incoherent particlelike behaviour, described by a Quantum 
Boltzmann equation, 
and a wavefunction description by means of Gross-Pitaevskii 
equation, 
is important for understanding Bose-Einstein 
condensation~\cite{zoller}. 
In~\cite{japan,berlin} it is shown 
how the formalism we have used in the present paper copes with
the more general problem of non-equilibrium macroscopic systems. 
However a systematic treatment of irreversibility in the very
similar problem of atomic interferometry involves QED
and is a future challenge.
  
\end{document}